\documentclass[12pt]{article}
\usepackage[left=2.5cm,right=2.5cm,top=2.5cm,bottom=2.5cm]{geometry}
\usepackage{amsfonts,amssymb}
\usepackage[utf8]{inputenc}
\usepackage{xcolor}
\usepackage{amsmath}
\usepackage{framed}
\colorlet{shadecolor}{lightgray}
\usepackage{hyperref}
\usepackage{cite}
\usepackage{parskip}

\begin{document} 

\baselineskip 18pt

\begin{center}

{\Large \bf Soft factors from classical scattering on the Reissner-Nordstr\"om spacetime}

\end{center}

\vskip .6cm
\medskip

\vspace*{2.0ex}

\baselineskip=18pt

\centerline{\large \rm Karan Fernandes$^{a}$ and Arpita Mitra$^{b}$}

\vspace*{4.0ex}

\centerline{\large \it ~$^a$Harish-Chandra Research Institute,}
\centerline{\large \it  Chhatnag Road, Jhusi, Prayagraj 211019, India.}
\vspace*{0.5ex}
\centerline{\large \it ~$^b$Indian Institute of Science Education \& Research Bhopal,}
\centerline{\large \it  Bhopal Bypass Road, Bhopal 420066, India.}

\vspace*{1.0ex}
\centerline{\small E-mail: karanfernandes@hri.res.in, arpitam@iiserb.ac.in}

\vspace*{5.0ex}

\centerline{\bf Abstract} \bigskip

We consider perturbations of the $4$ dimensional Reissner-Nordstr\"om spacetime induced by the probe scattering of a point particle with charge and mass moving on an unbound trajectory with an asymptotically large velocity. The resulting classical radiative solutions are the gravitational and electromagnetic bremmstrahlung. We use these classical solutions to derive the universal photon and graviton soft factor contributions at the tree level, which have the same form as noted in the literature. The soft factor expressions enable us to investigate the tail contribution to the memory effect in late time gravitational and electromagnetic waveforms. We find that generically, the contribution from the charge dominates that from the mass in the late time radiation.

\vfill \eject

\baselineskip 18pt

\tableofcontents

%%%%%%%%%%%%%%%%%%%%%%
\section{Introduction}
%%%%%%%%%%%%%%%%%%%%%%%

The radiation emitted by accelerated particles (bremsstrahlung) on curved spacetimes  cover an important class of scattering problems in classical General Relativity. Solutions include the electromagnetic radiation emitted by point charges~\cite{DeWitt:1960fc, Peters:1973ah,Westpfahl:1985tsl} and the gravitational radiation emitted by point masses~\cite{Peters:1970mx, Kovacs:1977uw} on the Schwarzschild background. 
One of the reasons for interest in these solutions concerns the power spectrum of the emitted radiation, which gives an estimate on their energy and angular distribution. An interesting regime in this spectrum lies in the soft limit, $\omega \to 0$. 
As this limit involves long wavelengths, the power spectrum result can be expected to be independent of specific details concerning the internal structure of the bodies and the nature of their scattering. This was demonstrated in~\cite{Smarr:1977fy}, where the zero-frequency limit of gravitational bremmstrahlung was considered in the case of distant encounters as well as head-on collisions. The soft limit result derived by Smarr at the linearized level agrees with post-linear results and the full non-linear numerical result for the head-on collision of equal mass BHs~\cite{Sperhake:2008ga, East:2012mb, Gruzinov:2014moa, Ciafaloni:2015xsr, Hopper:2017qus}

The soft limit is also of particular importance in the context of soft theorems on asymptotically flat spacetimes. Soft theorems relate scattering amplitudes involving soft particles with amplitudes without soft particles through a soft factor. The leading pole contribution to the soft photon and graviton factors~\cite{Weinberg:1964ew, Weinberg:1965nx} and soft photon expansions of the S-matrix~\cite{GellMann:1954kc,Low:1954kd,Burnett:1967km} pertained to infrared properties of quantum field theories. Soft theorems are now known to be more generally related with asymptotic symmetries on aymptotically flat spacetimes~\cite{Strominger:2013jfa,He:2014laa,He:2014cra,Lysov:2014csa,Campiglia:2014yka,Kapec:2014zla,Campiglia:2015yka,Avery:2015gxa,Strominger:2017zoo,Mao:2017tey,Laddha:2017vfh,Hamada:2018vrw,Ashtekar:2018lor}. The universal contributions in soft factors result as a consequence of invariance under $U(1)$ transformations in the case of abelian gauge theories and diffeomorphisms in the case of gravitational theories. There have been many recent developments of soft theorems concerning subleading contributions~\cite{Cachazo:2014fwa,Casali:2014xpa,Afkhami-Jeddi:2014fia,Schwab:2014xua,Broedel:2014fsa,Zlotnikov:2014sva,Kalousios:2014uva,Luo:2014wea,Conde:2016csj,DiVecchia:2016amo,Campiglia:2016efb}, loop corrections~\cite{Bern:2014vva,He:2014bga,Bianchi:2014gla,He:2017fsb,Sen:2017nim,Banerjee:2017jeg,Sahoo:2018lxl,DiVecchia:2018dob,Campiglia:2019wxe,DiVecchia:2019kle,AtulBhatkar:2019vcb} and multiple soft particles~\cite{Cachazo:2015ksa,Volovich:2015yoa,Klose:2015xoa,Georgiou:2015jfa,DiVecchia:2015bfa,Low:2015ogb,Saha:2016kjr,Saha:2017yqi,Chakrabarti:2017ltl,Zlotnikov:2017ahq,Chakrabarti:2017zmh,AtulBhatkar:2018kfi}. The universal contributions in soft factors and their connection with asymptotic symmetries have in particular led to several investigations concerning gravitational waves and the memory effect~\cite{Strominger:2014pwa,Pasterski:2015tva,Pasterski:2015zua,Mao:2017axa,Pate:2017vwa,Hamada:2017atr,Laddha:2018rle,Laddha:2018myi,Laddha:2018vbn}.
%   \\

This paper is motivated by the results of \cite{Laddha:2018rle}, where it was shown that universal contributions in the photon and graviton soft factors in any spacetime dimensions can be derived from classical electromagnetic and gravitational scattering processes. The use of classical scattering processes in determining soft factors is particularly relevant in four spacetime dimensions, where the S-matrix suffers from infrared divergences. In~\cite{Laddha:2018myi}, the soft factors were shown to develop logarithmic terms in the classical limit of soft theorems in four spacetime dimensions, arising from the long range interactions in electromagnetism and gravity. Their analysis also demonstrated that the classical gravitational bremmstrahlung from a point mass on the Schwarzschild spacetime~\cite{Peters:1970mx} satisfies the soft graviton theorem in four spacetime dimensions. Furthermore, the comparison with~\cite{Peters:1970mx} also revealed the presence of an overall phase, which influences the shape of the gravitational waveform. This additional phase does not follow from taking the classical limit of soft theorems and can only be inferred on comparison with classical results. The soft graviton factor involving the phase correction has been used to identify their effect on gravitational waves~\cite{Laddha:2018vbn}. In addition to the gravitational memory effect which follows from the leading contribution, the logarithmic terms present in the subleading contribution of the soft factor imply the presence of a tail term. The subleading contribution also involves the phase correction, which causes the tail term to vanish asymptotically in the case of either ultrarelativistic or massless particles.

Several approaches to perturbations on the Reissner-Nordstr\"om (RN) spacetime are known in the literature, notably those in~\cite{Zerilli:1974ai,Moncrief:1975sb,Chandrasekhar:1979iz}. However, our analysis in this paper will concern the weak field, fast-motion approximation which cannot be addressed using either the quadrupole moment formalism or the slow-motion approximation. The expansion in terms of generalized spherical harmonics is useful when the radiation is known to arise from specific multipole moments. The low frequency result then follows from taking the limit of the sum over moments, which in general is a more cumbersome route to soft limit expressions. For these reasons, in this paper we extend the result of~\cite{Peters:1970mx,Peters:1973ah} and consider the scattering of a point particle with mass and charge on the four dimensional RN spacetime, with the particle velocity taking on asymptotically large values. The scattering is considered in the probe limit with the gravitational and electromagnetic radiation results derived up to order $\frac{M}{r}$ and $\frac{Q}{r}$, where $M$ and $Q$ denote the mass and charge of the RN black hole. To the best of our knowledge, these classical solutions representing the gravitational and electromagnetic bremsstrahlung on the RN spacetime have not been previously derived in the literature.

We subsequently use our classical radiation results to demonstrate that they provide the expected tree level soft factor involved in the soft photon and graviton theorems, including phases. The property of soft factors involving an independent sum over incoming and outgoing particles implies that we can use the universal contributions to investigate classical observables, in broader settings involving interacting masses and charges. We use the soft factors to determine the memory effect and its tail contribution in late time waveforms, resulting from scattering processes involving several outgoing light particles and no incoming light particles. Our results demonstrate that the charge contribution in the tail generically dominates the mass. In considering the ultrarelativistic limit, both gravitational and electromagnetic wave tails vanish, consistent with the prior literature. 

Our paper is organized as follows. In the following section, we first review perturbations of charged black hole spacetimes before arriving at the equations on the linearized RN spacetime. The pertubative solutions in frequency space are then derived using the scalar Green function based on the worldline formalism on curved backgrounds. The derivation of the scalar Green function on linearized curved backgrounds has been provided in Appendix~\ref{gfrn} and specific expressions on the linearized RN spacetime have been derived in Appendix~\ref{app1}. In Sec.~\ref{soft}, we use the classical radiative solutions to determine the photon and graviton soft factors. We first review the results of~\cite{Laddha:2018rle,Laddha:2018myi,Sahoo:2018lxl}, which cover the predicted soft factors, the phase corrections involved and the integral identities needed to derive these expressions from classical radiative solutions. We then derive the photon soft factor in Sec.~\ref{electrosoft} and the graviton soft factor in Sec.~\ref{gravisoft}. 
Following~\cite{Laddha:2018vbn}, in Sec.~\ref{gravemwav} we determine the memory effect and its tail contribution in the late time gravitational and electromagnetic waveforms using the soft factor expressions. 
We conclude with a summary of our results.

%%%%%%%%%%%%%%%%%%%%%%%%%%%%%%%%%%
\section{Classical radiative solutions of probe scattering on the RN spacetime} \label{classical}
%%%%%%%%%%%%%%%%%%%%%%%%%%%%%%%%%%
In this section, we will derive the metric and electromagnetic perturbations of the RN spacetime due to the scattering of a massive and charged point particle travelling on an unbound trajectory with large impact parameter from the central black hole. We will follow the procedure used to calculate metric perturbations resulting from a point mass~\cite{Peters:1970mx} and electromagnetic perturbations due to a point charge~\cite{Peters:1973ah} on the Schwarzschild spacetime. In the following subsection, we consider linear perturbations of the Einstein-Maxwell equations of a charged black hole spacetimes due to point particle with mass and charge. We then specifically consider these perturbations on the RN spacetime up to linear order in $\displaystyle{\frac{M}{r}}$ and $\displaystyle{\frac{Q}{r}}$, in accordance with our assumption of a large impact parameter. The perturbations are then solved in the last subsection in frequency space by using the worldline formalism for the scalar Green function.

%%%%%%%%%%%%%%%%%%%%%%%%%%%%%%%%%%%%%%%%%%%%%
\subsection{Perturbations of Einstein-Maxwell equations}
%%%%%%%%%%%%%%%%%%%%%%%%%%%%%%%%%%%%%%%%%%%
We consider the background to be a charged black hole spacetime which is a solution of the Einstein-Maxwell action. The solution satisfies Einsteins equations
\begin{equation}
G_{\mu\nu} - T_{\mu\nu} = 0\;\label{ein.eq}
\end{equation}
and the source-free Maxwell equations
\begin{equation}
F^{\mu\nu};_{\nu}=0 \label{max.eq}
\end{equation}
where 
\begin{align*}
G_{\mu\nu} =R_{\mu\nu}-\frac{R}{2}g_{\mu\nu}\;, \quad F_{\mu\nu} =A_{\nu},_{\mu}-A_{\mu},_{\nu}\;, \quad T_{\mu\nu} =\frac{1}{4\pi}\left(F_{\mu\alpha}F_{\nu\beta}g^{\alpha\beta}-\frac{1}{4}g_{\mu\nu}F_{\alpha\beta}F_{\gamma\delta}g^{\alpha\gamma}g^{\beta\delta}\right) \,. 
\end{align*}
The background metric and gauge potential are $g_{\mu\nu}$ and $A_{\mu}$. We adopt the mostly plus convention for the metric and set $8 \pi G = 1$. Semicolons indicate covariant derivatives, while commas denote partial derivatives. Our convention for the Riemann tensor is
\begin{equation}
R^{\alpha}_{\phantom{\alpha}\mu \beta \nu}  = \Gamma^{\alpha}_{\mu \nu},_{\beta} - \Gamma^{\alpha}_{\mu \beta},_{\nu} + \Gamma^{\alpha}_{\beta \gamma}\Gamma^{\gamma}_{\mu \nu} - \Gamma^{\alpha}_{\nu \gamma} \Gamma^{\gamma}_{\mu \beta} \,,
\end{equation}
where 
\begin{equation}
\Gamma^{\alpha}_{\mu\nu} = \frac{1}{2} g^{\alpha \beta} \left(g_{\beta \nu},_{\mu} + g_{\beta \mu},_{\nu} - g_{\mu\nu},_{\beta}\right) 
\end{equation}

We now linearly perturb the spacetime by introducing a point particle with mass and charge. The corresponding action is
\begin{equation}
S_{P} = -m \int d\sigma \sqrt{-g_{\mu\nu}\frac{dr^{\mu}}{d \sigma}\frac{dr^{\nu}}{d \sigma}} + \frac{q}{4\pi} \int d\sigma A_{\mu}\frac{dr^{\mu}}{d \sigma}\,.
\label{act.pp}
\end{equation}
From the variation of this action, we derive the following stress tensor $T^{\mu \nu}_{(P)}$ and current $J^{\mu}_{(P)}$
\begin{align}
T^{\mu \nu}_{(P)} &=m\int \delta(x,r(\sigma))\frac{dr^{\mu}}{d \sigma}\frac{dr^{\nu}}{d \sigma}\,d \sigma \,, \notag \\
J_{(P)}^{\mu}&= \frac{q}{4\pi} \int \delta(x,r(\sigma))\frac{dr^{\mu}}{d\sigma} \, d \sigma
\end{align}
where $\delta(x,r(\sigma))$ refers to the covariant delta function which is normalized as
\begin{equation}
\int \sqrt{-g}~\delta(x,r(\sigma)) d\sigma =1
\end{equation}
The stress-energy tensor and current of the point particle induces a perturbation of the metric and gauge potential which we denote by $h_{\mu \nu}$ and $a_{\mu}$ respectively, i.e.
\begin{align}
g_{\mu\nu} &\to g_{\mu\nu}+ \delta g_{\mu\nu} = g_{\mu\nu} + 2 h_{\mu \nu}\,, \notag\\
A_{\mu} &\to A_{\mu}+ \delta A_{\mu} = A_{\mu} + a_{\mu}\,.
\end{align}
The variation of Eq.~(\ref{ein.eq}) and Eq.~(\ref{max.eq}) provides the following equations
\begin{align}
\delta G_{\mu\nu}  - \delta T^{h}_{\mu\nu} - \delta T^{a}_{\mu\nu} &=  T_{\mu\nu}^{(P)}\label{ein.pert}\\
\delta(F^{\mu\nu};_{\nu})&=  4\pi J^{\mu}_{(P)} \,,\label{max.pert}
\end{align}
where we have denoted the perturbation of the stress-energy tensor as a combination of terms involving perturbations of the metric $\delta T^{h}_{\mu\nu}$ and perturbations of the gauge potential $\delta T^{a}_{\mu\nu}$ in Eq.~(\ref{ein.pert}).

We will denote the perturbed electromagnetic field strength tensor by 
\begin{equation}
f_{\mu\nu}= a_{\nu,\mu} - a_{\mu,\nu} \,.
\end{equation}
It will also be convenient to describe the perturbation equations in terms of trace-reversed metric perturbations $e_{\mu \nu}$ defined by
\begin{equation} 
e_{\mu\nu}=h_{\mu\nu}-\frac{1}{2}h g_{\mu\nu} \,; \qquad  h = g^{\mu \nu} h_{\mu \nu} = - g^{\mu \nu} e_{\mu \nu} = - e
\end{equation}

The simplified expressions for $\delta G_{\mu\nu} \,, \delta T^{h}_{\mu \nu}\,, \delta T^{a}_{\mu\nu}$ and $\delta(F^{\mu\nu};_{\nu})$ in terms of $e_{\mu \nu}$ and $f_{\mu\nu}$ are
\begin{align}
\delta G_{\mu\nu}&= -e_{\mu\nu;\alpha}{}^{\alpha} + e_{\mu\alpha;}{}^{\alpha}{}_{\nu} + e_{\nu\alpha;\phantom{\alpha}\mu}^{\phantom{\nu\alpha;}\alpha}+ \left(R_{\nu}{}^{\delta}e_{\delta\mu}+R_{\mu}{}^{\delta}e_{\delta\nu}\right)+2R^{\alpha}{}_{\nu\mu}{}^{\delta}e_{\delta\alpha}  \notag\\&  \qquad \qquad -  g_{\mu\nu} e_{\alpha\beta;}{}^{\alpha\beta} - R e_{\mu\nu}+ g_{\mu\nu} R^{\alpha\beta} e_{\alpha\beta}\notag\\
\delta T^{h}_{\mu\nu} &= e T_{\mu \nu} -\frac{1}{8 \pi}e_{\mu\nu} F_{\alpha \beta} F^{\alpha \beta} - \frac{1}{2 \pi}g^{\alpha \epsilon}g^{\beta \delta}e_{\epsilon\delta}\left( F_{\alpha \mu} F_{\beta \nu} - \frac{1}{4} g_{\mu\nu} F_{\alpha \gamma} F_{\beta}{}^{\gamma}\right)\notag\\
\delta T^{a}_{\mu\nu} &= \frac{1}{4 \pi}g^{\alpha \beta}\left( f_{\alpha \mu} F_{\beta \nu} + f_{\alpha \nu} F_{\beta \mu} - \frac{1}{2}g_{\mu\nu}g^{\gamma \delta} f_{\alpha \gamma} F_{\beta \delta}\right)\notag\\
\delta(F^{\mu\nu};_{\nu}) & = - g^{\alpha \rho} g^{\mu \nu}\left(2 g^{\beta \sigma} e_{\rho \sigma}F_{\nu \beta};_{\alpha} - f_{\nu \rho};_{\alpha} - 2 F_{\alpha \nu} e_{\rho \beta;}{}^{\beta} \right)
\label{del.exp}
\end{align}

By substituting the first three expressions of Eq.~(\ref{del.exp}) in Eq.~(\ref{ein.pert}), we find the following expression for the perturbed Einstein equation 
\begin{align}
-T^{(P)}_{\mu \nu} &= e_{\mu\nu;\alpha}{}^{\alpha} - e_{\mu\alpha;}{}^{\alpha}{}_{\nu} - e_{\nu\alpha;\phantom{\alpha}\mu}^{\phantom{\nu\alpha;}\alpha} - \left(R_{\nu}{}^{\delta}e_{\delta\mu}+R_{\mu}{}^{\delta}e_{\delta\nu}\right) - 2R^{\alpha}{}_{\nu\mu}{}^{\delta}e_{\delta\alpha} + R e_{\mu\nu} \notag\\& \quad - g_{\mu\nu} R^{\alpha\beta} e_{\alpha\beta} + g_{\mu\nu} e_{\alpha\beta;}{}^{\alpha\beta} + \frac{1}{4\pi} g^{\alpha \beta}\left( f_{\alpha \mu} F_{\beta \nu} + f_{\alpha \nu} F_{\beta \mu} - \frac{1}{2}g_{\mu\nu}g^{\gamma \delta} f_{\alpha \gamma} F_{\beta \delta}\right)\notag\\ & \qquad +  e T_{\mu \nu} - \frac{1}{8\pi}e_{\mu\nu} F_{\alpha \beta} F^{\alpha \beta} - \frac{1}{2\pi} g^{\alpha \epsilon}g^{\beta \delta}e_{\epsilon\delta}\left( F_{\alpha \mu} F_{\beta \nu} - \frac{1}{4} g_{\mu\nu} F_{\alpha \gamma} F_{\beta}{}^{\gamma}\right)\,,
\label{ein.pertha}
\end{align}
while using the last line of Eq.~(\ref{del.exp}) in Eq.~(\ref{max.pert}) gives us the perturbed Maxwell equation
\begin{equation}
-4\pi g^{\mu \nu} J_{\nu}^{(P)}= g^{\alpha \rho} g^{\mu \nu}\left(2 g^{\beta \sigma} e_{\rho \sigma}F_{\nu \beta};_{\alpha} - f_{\nu \rho};_{\alpha} - 2 F_{\alpha \nu} e_{\rho \beta;}{}^{\beta} \right)
\label{max.pertha}
\end{equation}

In general, the conservation equation requires
\begin{equation}
\delta G_{\mu\nu;}{}^{\mu} - \delta T^{h}_{\mu\nu;}{}^{\mu} - \delta T^{a}_{\mu\nu;}{}^{\mu} -T^{(P)}_{\mu\nu;}{}^{\mu} = 0
\end{equation} 

%%%%%%%%%%%%%%%%%%%%%%%
\subsection{Perturbations on the linearized RN spacetime}
%%%%%%%%%%%%%%%%%%%%%%%%%

We will now consider the above perturbations on the RN spacetime. In isotropic coordinates, the metric is given by
\begin{equation}
ds^2=-g_{00}dt^2+g_{ij}dx^i dx^j \label{rn.met}
\end{equation}
with
\begin{align}
g_{00} =-\left(\frac{1-\frac{1}{4R^2}\left(\frac{M^2}{64 \pi^2} - \frac{Q^2}{8\pi}\right)}{\left(1+\frac{M}{16 \pi R}\right)^2-\frac{1}{8\pi} \left(\frac{Q}{2R}\right)^2}\right)^2\,, \qquad g_{ij} = \delta_{ij} \left(\left(1+\frac{M}{16 \pi R}\right)^2- \frac{1}{8\pi} \left(\frac{Q}{2R}\right)^2\right)^2 \,.
\label{rn.metc}
\end{align}
where $R = \vert \vec{x} \vert$. The parameters $M$ and $Q$ refer to the mass and charge, respectively, of the RN black hole. The gauge field $A_{\mu}$ has the following non-vanishing component 
\begin{equation}
A_0=\frac{Q}{R}\left(1+\frac{M}{8\pi R}+\frac{1}{4R^2}\left(\frac{M^2}{64 \pi^2} - \frac{Q^2}{8 \pi}\right)\right)^{-1} \label{rn.gf}
\end{equation}
Eq.~(\ref{rn.met}) and Eq.~(\ref{rn.gf}) satisfy the Einstein-Maxwell equations in Eq.~(\ref{ein.eq}) and Eq.~(\ref{max.eq}). 
In considering perturbations about this background along the lines of the previous section, we will implement a few approximations. First, we will be working in the probe scattering limit, where the point particle is assumed to have a large impact parameter from the central black hole. This amounts to the assumption that we consider $\frac{M}{R} << 1$ and $\frac{Q}{R} <<1$. Second, the mass $m$ and charge $q$ of the point particle is considered to be much smaller than the mass $M$ and charge $Q$ of the RN black hole. Specifically, we make the assumption that $M >> m$ and $Q >> q$. These assumptions allow us to consider our perturbations about the linearized RN spacetime, where the metric and gauge potential will be considered only up to terms which are linear in $Q$ and $M$. 
We will express the metric and gauge field in terms of the potential
\begin{equation}
\phi\left(\vec{x}\right)= -\frac{M}{8 \pi R}\,.
\label{rn.pot}
\end{equation}
Hence all $\mathcal{O}\left(\phi^2\right)$ contributions will contain the $\mathcal{O}\left(M^2\right)\,, \mathcal{O}\left(Q M\right)$ and $\mathcal{O}\left(Q^2\right)$ terms. 
Ignoring $\mathcal{O}\left(\phi^2\right) $ corrections, the metric components then have the form
\begin{align}
g_{00} = -\left(1 + 2 \phi\right) \,, \qquad g_{00} = 0 \,, \qquad g_{ij} = \delta_{ij} \left(1 - 2\phi\right)\,,
\label{rn.meta}
\end{align}
while the gauge potential in this approximation is
\begin{equation}
A_0(\vec{x}) = -\frac{8 \pi Q}{M}\phi\left(\vec{x}\right)  \label{rn.gfa}
\end{equation}
We will denote $\frac{\partial}{\partial x^i}$ as $\partial_i$ and $\frac{\partial}{\partial x^0}$ as $\partial_0$. We then have from Eq.~(\ref{rn.meta}) the following non-vanishing connection, Riemann and Ricci tensor components  
\begin{align}
\Gamma^{0}_{0i} = \phi_{,i} = \Gamma^{i}_{00} \,, \qquad &\Gamma^{i}_{jk} = \delta_{jk}\phi_{,i} - \delta_{ij}\phi_{,k} - \delta_{ik}\phi_{,j}\,, \notag\\
R^{0}{}_{ij0} = \phi_{,ij} \,, \qquad  &R^k{}_{ijm} = \delta_{im}\phi_{,kj} + \delta_{jk}\phi_{,im} - \delta_{ij}\phi_{,mk} - \delta_{mk}\phi_{,ij} \notag\\
R_{00} = \phi_{,kk} \,, \qquad &R_{ij} = \delta_{ij} \phi_{,kk} \,, \notag\\
R=&2 \phi_{,kk}
\label{met.comp}
\end{align}
while Eq.~(\ref{rn.gfa}) gives us the non-vanishing electromagnetic field strength tensor component
\begin{equation}
F_{0i} = \frac{8 \pi Q}{M} \phi_{,i} \,.
\label{gf.comp}
\end{equation}
At the linearized level, the stress-energy tensor of the Maxwell field goes like $\mathcal{O}\left(\phi^2\right)$. Indeed, this also implies that the linearized RN spacetime appears to be exactly as that of linearized Schwarzschild spacetime with a gauge potential. More significantly however, perturbations about the linearized RN spacetime are different from those about the linearized Schwarzschild spacetime, which will be relevant in the following analysis. Thus while $\delta T^h_{\mu\nu}$ vanishes at the linearized level, contributions from $\delta T^a_{\mu\nu}$ are present in the perturbed Einstein equation in Eq.~(\ref{ein.pertha}). Likewise, the perturbed Maxwell equation in Eq.~(\ref{max.pertha}) involve $e_{\mu\nu}$ corrections unlike the perturbations about the Schwarzschild spacetime. 

We will now substitute Eq.~(\ref{rn.met}) - Eq.~(\ref{gf.comp}) in Eq.~(\ref{ein.pertha}) and Eq.~(\ref{max.pertha}). Let us define
\begin{equation}
\Box = \eta^{\mu \nu} \partial_{\mu}\partial_{\nu} = -\partial^2_0+ \partial^2_i 
\label{box}
\end{equation}
and
\begin{equation}
k_{\mu} = e_{\mu\nu};{}^{\nu} \,, \qquad l = -a_{0,0} + a_{i,i}\,.
\label{gauge}
\end{equation}
We will be interested in the radiative components of the gravitational and electromagnetic fields. These components will only require us to consider the spatial components of the perturbed fields. As we will note, other components of the perturbed fields can be determined from the gauge fixing choices we will adopt. For the $ij$ component of the perturbed Einstein equation, Eq.~(\ref{ein.pertha}), we find the following expression
\begin{align}
-T^{(P)}_{i j} &= \Box\left(\left(1+2\phi\right) e_{ij}\right) - k_{i,j} - k_{j,i} - 2\left(\phi_{,k}e_{ki}\right)_{,j} - 2\left(\phi_{,k}e_{kj}\right)_{,i} - \left(k_{0,0} - k_{l,l}\right) \delta_{ij}  \notag\\& \quad +4 \left[ \phi e_{ij,00} + \phi_{,i} e_{0j,0}+ \phi_{,j} e_{i0,0} +\frac{1}{2}\left(\phi_{,ij} - \frac{1}{2}\phi_{kk}\delta_{ij}\right) \left(e_{00}+ e_{ll}\right)\right]\notag\\ & \qquad + 2 \delta_{ij}\left(\phi_{,kl} e_{kl} - \phi_{,k}k_{k}+ 2 \phi k_{0,0}\right) - \frac{2 Q}{M}\left(f_{0i}\phi_{,j}+f_{0j}\phi_{,i} - \delta_{ij}f_{0l}\phi_{,l}\right)\,,
\label{ein.pertf}
\end{align}
while the $i$ component of the perturbed Maxwell equation, Eq.~(\ref{max.pertha}), gives us
\begin{align}
-\left(1+2\phi\right)4\pi J_{i}^{(P)}&= \Box a_i - l_{,i} + 2\phi_{,k}f_{ki} + 4\phi\left(a_{i,kk} - a_{k,ki}\right) +\frac{16 \pi Q}{M}\left(e_{0j}\phi_{,ij} + \phi_{,i}k_0\right) \label{ai.pertf}
\end{align}
We now need to fix a gauge in Eq.~(\ref{ein.pertf}) and Eq.~(\ref{ai.pertf}). It is desirable that any choice we make for $k_{\mu}$ and $l$ reduce to the usual flat spacetime de Donder and Lorenz gauges in the $\phi \to 0$ limit. However, the key principle in our choice will be to ensure that all mixed terms with $\phi$ involve only time derivatives of the perturbed fields. This will be crucial for solving the perturbations using the approach described in the next subsection.

Thus in Eq.~(\ref{ein.pertf}), we need to implement a choice for $k_\mu$ such that it eliminates the terms
\begin{equation}  
- 2\left(\left(\phi_{,k}e_{ki}\right)_{,j} + \left(\phi_{,k}e_{kj}\right)_{,i} - \delta_{ij}\phi_{,kl} e_{kl}\right) + \frac{2Q}{M}\left(a_{0,i}\phi_{,j} + a_{0,j}\phi_{,i} -\delta_{ij} a_{0,l}\phi_{,l}\right)
\end{equation}
We note that the following gauge choice for $k_{\mu}$ achieves this
\begin{equation}
k_{\mu} = -2\phi_{,k}e_{k\mu} + \frac{2Q}{M} a_0 \phi_{,\mu} \,.
\label{k.exp}
\end{equation}
Substituting this expression for $k_{\mu}$ in Eq.~(\ref{ein.pertf}) gives us the desired gauge fixed expression
\begin{align}
-T^{(P)}_{i j} &= \Box\left(\left(1+2\phi\right) e_{ij}\right) +4 \left[ \phi e_{ij,00} + \phi_{,i} e_{0j,0}+ \phi_{,j} e_{i0,0} +\frac{1}{2}\left(\phi_{,ij} - \frac{1}{2}\phi_{kk}\delta_{ij}\right) \left(e_{00}+ e_{ll}\right)\right]\notag\\ & \quad  - \frac{2 Q}{M}\left[a_{i,0}\phi_{,j}+a_{j,0}\phi_{,i} - \delta_{ij}a_{l,0}\phi_{,l} + 2 a_0\left(\phi_{,ij} - \frac{1}{2}\delta_{ij}\phi_{kk}\right)\right]\,.
\label{ein.pertfg}
\end{align}
Substituting Eq.~(\ref{k.exp}) in Eq.~(\ref{ai.pertf}) also eliminates the $k_{\mu}$ contribution in the electromagnetic perturbation equation, since these contributions are $\mathcal{O}\left(\phi^2\right)$. Thus the only terms which need to be addressed in the electromagnetic pertubation equation are
\begin{equation}
2\phi_{,k}f_{ki} + 4\phi\left(a_{i,kk} - a_{k,ki}\right)
\end{equation} 
This can be achieved by the following choice for $l$
\begin{equation}
l = -4 \phi a_{k,k} - 2 \phi_{,k}a_k
\label{l.exp}
\end{equation}
We now substitute Eq.~(\ref{l.exp}) in Eq.~(\ref{ai.pertf}) to find the gauge fixed expression 
\begin{equation}
-\left(1 - \phi\right) 4 \pi J^{(P)}_i = \Box\left(\left(1+\phi\right)a_i\right) + 4\left[\phi a_{i,00}  + \phi_{,i} a_{0,0}\right] - \phi_{,kk}a_i + 2\phi_{,ik}a_k + \frac{16 \pi Q}{M}e_{0j}\phi_{,ij} \,,
\label{ai.pertfg}
\end{equation}
where we made use of the following property resulting from Eq.~(\ref{l.exp})
\begin{equation}
\phi l = 0 = \phi \left(-a_{0,0} + a_{k,k}\right)
\end{equation}

In the following subsections, we will proceed to solve the equations for $e_{ij}$ and $a_i$ given in Eq.~(\ref{ein.pertfg}) and Eq.~(\ref{ai.pertfg}) respectively. These solutions will be determined in frequency space following a Fourier transform in time.
The procedure which will be used to derive the solutions is the worldline formalism applied to the scalar Green function.

%%%%%%%%%%%%%%%%%%%%%%%%%%%%%%
\subsection{Solutions using the worldline formalism}
%%%%%%%%%%%%%%%%%%%%%%%%%%%%%%%%%
Eq.~(\ref{ein.pertfg}) and Eq.~(\ref{ai.pertfg}) can be viewed as differential equations involving the flat spacetime D'Alembertian operator and subleading terms involving $\phi$ and its derivatives. To obtain the solutions of such equations in frequency space 
we will need the solution of the following equation
\begin{equation}
\psi^{(1)}_{\phantom{(1)};\alpha}{}^{\alpha} = -\int \delta(x,r(\sigma)) f(\sigma) d\sigma + \mathcal{O}\left(R^2\right)\,,
\label{scalar.cov}
\end{equation}
with the form of $f(\sigma)$ depending on the source and where the covariant delta function $\delta(x,r(\sigma))$ is related to the flat spacetime delta function $\delta^{4}\left(x-r(\sigma)\right)$ via
\begin{equation}
\delta(x,r(\sigma)) \sqrt{-g} = \delta^{4}\left(x-r(\sigma)\right) = \delta(t - r^0(\sigma)) \delta^{(3)}(\vec{x} - \vec{r}(\sigma))\,.
\end{equation}
The solution of Eq.~(\ref{scalar.cov}) which can be derived using the worldline formalism~\cite{J.L.Synge:1960zz, Peters:1966} is \footnote{A brief review of this derivation has been provided in Appendix \ref{gfrn}}
\begin{align}
\psi^{(1)} = \psi^{(0)} + \delta \psi^{(0)}\,, 
\label{psi1.sol}
\end{align}
where
\begin{align}
\psi^{(0)}(x) &=  \frac{1}{4\pi}\int \limits_{-\infty}^{\sigma_0} \delta\left(-\Omega\left(x,r(\sigma)\right)\right) f(\sigma) d\sigma \,,
\label{psi0.sol} \\
\delta \psi^{(0)}(x) &= \frac{1}{16 \pi^2} \int \sqrt{-g(y)} \delta\left(-\Omega\left(x,y\right)\right) d^4y  \int \limits_{-\infty}^{\sigma_0} \delta'\left(-\Omega\left(y,r(\sigma)\right)\right) F\left(y,r(\sigma)\right) f(\sigma) d\sigma\,. \label{dpsi.sol}
\end{align}
In Eq.~(\ref{psi0.sol}), the world function $\Omega\left(x,r(\sigma)\right)$ is defined in terms of the geodesic $U^{\mu}$ which connects the point $x$ at parametric value $u_1$ and $r(\sigma)$ at parametric value $u_0$ 
\begin{equation}
\Omega\left(x,r(\sigma)\right) = \frac{u_1 - u_0}{2} \int \limits_{u_0}^{u_1}  g_{\mu \nu}U^{\mu} U^{\nu}\, du \,,
\label{wf.sol}
\end{equation}
while $F\left(y ,r(\sigma)\right)$ in Eq.~(\ref{dpsi.sol}) is further defined using the Ricci tensor $R_{\mu\nu}$
\begin{equation}
F\left(y ,r(\sigma)\right) = \frac{1}{u_1 - u_0} \int \limits_{u_0}^{u_1} \left(u-u_0\right)^2 R_{\mu \nu} U^{\mu} U^{\nu} du\,.
\label{f.sol}
\end{equation}

We note at this point that that while $\psi^{(0)}$ does involve terms that are both independent and linear in curvature, $\delta \psi^{(0)}(x)$ only involves curvature dependent terms. Let us now consider the specific case of the linearized RN spacetime, whose metric is given in Eq.~(\ref{rn.meta}). 
By expanding Eq.~(\ref{scalar.cov}) explicitly in terms of $\phi$, we find
\begin{equation}
\Box \psi^{(1)} + 4 \phi \partial_t^2 \psi^{(1)} = - \int \delta^{4}\left(x - r(\sigma)\right) f(\sigma) d\sigma + \mathcal{O}\left(R^2\right)\,.
\label{scalar.fe}
\end{equation}
The source term in Eq.~(\ref{scalar.fe}) is now simply that for the Green function in flat spacetime. This suggests that we can assume $\psi^{(0)}$ to be composed in the following way 
\begin{equation}
\psi^{(0)} = \psi_0^{(0)} + \psi_1^{(0)} \,.
\end{equation}
where $\psi_0^{(0)}$ manifestly satisfies the flat spacetime equation for the Green function, while $\psi_1^{(0)}$ is an additional contribution which does not. From Eq.~(\ref{scalar.fe}), we then find the following two equalities
\begin{align}
\Box  \psi_0^{(0)} &= - \int \delta^{4}\left(x - r(\sigma)\right) f(\sigma) d\sigma\,, \label{sceq.1}\\
-4 \phi \partial_t^2 \psi^{(0)}  &= \Box  \left(\psi_1^{(0)} + \delta \psi^{(0)}\right) \label{sceq.2}
\end{align}
Eq.~(\ref{sceq.1}) and Eq.~(\ref{sceq.2}) further imply that in considering perturbative solutions about the linearized RN spacetime, we can simply express Eq.~(\ref{scalar.fe}) as
\begin{equation}
\Box \left(\psi_0^{(0)} - \psi_1^{(0)} - \delta \psi^{(0)}\right)  = - \int \delta^{4}\left(x - r(\sigma)\right) f(\sigma) d\sigma + \mathcal{O}\left(R^2\right)\,. \label{sceq.fin}
\end{equation}
Hence on the linearized RN spacetime, we can express Eq.~(\ref{scalar.fe}) as an equation for the scalar Green function on flat spacetime. Furthermore, we can also derive solutions when we consider specific corrections of Eq.~(\ref{scalar.fe}), as in the case of Eq.~(\ref{ein.pertfg}) and Eq.~(\ref{ai.pertfg}). Should Eq.~(\ref{scalar.fe}) contain correction terms involving derivatives of $\phi$ and time derivatives of $\psi^{(1)}$, then Eq.~(\ref{sceq.2}) can be used to find an expression involving the flat spacetime D'Alembertian operator acting on the additional terms. We may thus perturbatively construct solutions in these cases as well.
  
In Appendix \ref{app1}, we have explicitly derived the following solutions for $\psi^{(0)}$ and $\delta \psi^{(0)}$ on the linearized RN spacetime
\begin{align}
\psi^{(0)}(t\,,\vec{x}) & = \frac{1}{4 \pi} \int \limits_{-\infty} ^{\infty} \frac{\delta\left(t - r^0 - R_0 -  \frac{M}{4\pi} \Gamma\left(\vec{x},\vec{z}\right) \right)}{R_0} f(\sigma) d\sigma \,, \label{psiRN0.sol} \\
\delta \psi^{(0)}(t\,,\vec{x}) & = \frac{M}{16 \pi^2}\partial_t \int \limits_{0}^{\infty} dv \int \limits_{-\infty}^{\infty} d\sigma \frac{\delta\left(t - r^0 - \vert \vec{r} \vert - v - \rho(v)\right)}{\rho(v) \left(\vert \vec{r}\vert + v\right)} f(\sigma) \,, \label{dp.sol}
\end{align}
where as before $\vec{R}_0 = \vec{x} - \vec{r}$, $R = \vert \vec{x}\vert$ and 
\begin{equation}
 \Gamma\left(\vec{x},\vec{r}\right) = \text{ln} \left(\frac{R R_0 + \vec{x}.\vec{r} }{\vert\vec{r}\vert R_0 + \vec{z}.\vec{r}}\right) \,, \qquad \rho(v) = \sqrt{R^2 + v^2 + \frac{2 v \vec{x}.\vec{r}}{\vert \vec{r}\vert}}
\label{gamrho.exp}
\end{equation}
The solutions in Eq.~(\ref{psiRN0.sol}) and Eq.~(\ref{dp.sol}) are the same as those about the linearized Schwarzschild spacetime~\cite{Peters:1966}.

To derive the perturbative solutions in frequency space, we perform the Fourier transform
\begin{align}
\tilde{\psi} \left(\omega\,, \vec{x}\right) &= \frac{1}{2\pi} \int dt\, e^{i \omega t}\, \psi \left(t, \vec{x}\right) \,, \label{ft.t}\\
\psi \left(t, \vec{x}\right) &= \int d \omega\, e^{-i \omega t} \,\tilde{\psi} \left(\omega\,, \vec{x}\right) \,, \label{ft.w}
\end{align}
where $\psi$ refers to any of the fields mentioned above $\psi^{(1)}\,, \psi^{(0)}_0 \,, \psi^{(0)}_1$ or $\delta \psi^{(0)}$.
By Fourier transforming Eq.~(\ref{psiRN0.sol}) and Eq.~(\ref{dp.sol}) (using Eq.~(\ref{ft.t})), we find
\begin{align}
\tilde{\psi}^{(0)}(\omega\,, \vec{x}) & = \frac{1}{4 \pi} \int \limits_{-\infty} ^{\infty} \frac{e^{i \omega\left(r^0 + R_0\right)}}{R_0} f(\sigma) d\sigma + \frac{i \omega M}{16 \pi^2}\int \limits_{-\infty} ^{\infty} \frac{e^{i \omega\left(r^0 + R_0\right)} \Gamma\left(\vec{x},\vec{r}\right)}{R_0}f(\sigma) d\sigma\,, \label{psi0.solf} \\
\delta \tilde{\psi}^{(0)}(\omega\,, \vec{x}) & = -\frac{i \omega M}{16 \pi^2} \int \limits_{0}^{\infty} du \int \limits_{-\infty}^{\infty} d\sigma \frac{e^{i \omega\left(r^0 + u + \vert\vec{r}\vert + \rho(u)\right)}}{\left(u+\vert \vec{r}\vert\right)\rho(u)} f(\sigma) \,, \label{dp.solf}
\end{align}
From Eq.~(\ref{psi0.solf}), we further identify
\begin{align}
\tilde{\psi}^{(0)}_0\left(\omega\,, \vec{x}\right) &= \frac{1}{4 \pi} \int \limits_{-\infty} ^{\infty} \frac{e^{i \omega\left(r^0 + R_0\right)}}{R_0} f(\sigma) d\sigma \,,\label{psi00.solf}\\
 \tilde{\psi}^{(0)}_1 \left(\omega\,, \vec{x}\right) &= \frac{i \omega M}{16 \pi^2}\int \limits_{-\infty} ^{\infty} \frac{e^{i \omega\left(r^0 + R_0\right)} \Gamma\left(\vec{x},\vec{r}\right)}{R_0} f(\sigma) d\sigma \label{psi01.solf}
\end{align}
On substituting Eq.~(\ref{ft.w}) in Eq.~(\ref{sceq.1}) and defining $\widetilde \Box := \omega^2 + \partial_i^2$, we find
\begin{equation}
\int d \omega \,e^{-i \omega t}\, \widetilde \Box \tilde{\psi}^{(0)}_0\left(\vec{x},\omega\right) = - \int \delta(t - r^0) \delta^{(3)}\left(\vec{x} - \vec{r}\right) f(\sigma) d\sigma \,,
\label{psi0.ftw}
\end{equation}
which is satisfied by the solution in Eq.~(\ref{psi00.solf}). By Fourier transforming Eq.~(\ref{sceq.2}) and on using Eq.~(\ref{psi01.solf}) and Eq.~(\ref{dp.solf}) we find
\begin{equation}
-\widetilde \Box G_{M}\left(\omega\,,\vec{x}\,,\vec{r}\right) = - \left(\omega^2 + \partial^2_i\right) G_{M}\left(\omega\,,\vec{x}\,,\vec{r}\right) = \phi\left(\vec{x}\right)\frac{e^{i\omega R_0}}{R_0} \,,
\label{gf.rn}
\end{equation}
where we have defined $G_{M}$ as
\begin{equation}
G_{M}\left(\omega\,,\vec{x}\,,\vec{r}\right) = -\frac{i M}{16 \pi \omega} \left(\frac{e^{i \omega R_0} \Gamma\left(\vec{x},\vec{r}\right)}{R_0} - \int \limits_{0}^{\infty} du \frac{e^{i \omega\left(u + \vert\vec{r}\vert + \rho(u)\right)}}{\left(u+\vert \vec{r}\vert\right)\rho(u)} \right)
\label{gfn}
\end{equation}
Eq.~(\ref{gf.rn}) enables us to express $\phi$ and its derivatives in terms of $\widetilde \Box G_{M}$. Let us define the derivative $\displaystyle{\nabla_{i} := \frac{\partial}{\partial x^i} + \frac{\partial}{\partial r^i}}$. Since $\vec{R}_0 = \vec{x} - \vec{r}$ we have $\nabla_i f(R_0) = 0$, for $f(R_0)$ being some function of $R_0$. Hence we can conveniently describe derivatives of $\phi$ in the following way
\begin{equation}
-\widetilde \Box \nabla_i G_{M}\left(\omega\,,\vec{x}\,,\vec{r}\right) = \phi_{,i}\frac{e^{i\omega R_0}}{R_0} \,, \quad  -\widetilde \Box \nabla_i \nabla_k G_{M}\left(\omega\,,\vec{x}\,,\vec{r}\right) = \phi_{,ik} \frac{e^{i\omega R_0}}{R_0} \,, \quad \text{etc.}
\label{gf.rn2}
\end{equation}

%%%%%%%%%%%%%%%%%%%%%%%%%%
\subsection{Solutions of the perturbation equations}
%%%%%%%%%%%%%%%%%%%%%%%%%%%%
Using the results of the previous subsection, we can now derive the solutions of Eq.~(\ref{ein.pertfg}) and Eq.~(\ref{ai.pertfg}). These equations can be viewed as involving corrections of the equation for the scalar Green function. In the case of Eq.~(\ref{ein.pertfg}), we have
\begin{equation}
 -T^{(P)}_{i j} =  -m\int \delta_{ki}\delta_{lj}\frac{\delta^4(x - r(\sigma))}{1 + 2\phi\left(\vec{r}\right)}\frac{dr^k}{d \sigma}\frac{dr^l}{d \sigma}\,d \sigma = \Box\left(\left(1+2\phi\right) e_{ij}\right) +4 \phi e_{ij,00} + \cdots\,,
\label{ein.pergt}
\end{equation}
while for Eq.~(\ref{ai.pertfg}) we have
\begin{equation}
-\left(1 - \phi\right) 4 \pi J^{(P)}_i = -q\int \delta_{ki}\frac{\delta^4(x - r(\sigma))}{1 + \phi\left(\vec{r}\right)}\frac{dr^k}{d \sigma}\,d \sigma = \Box\left(\left(1+\phi\right)a_i\right) + 4 \phi a_{i,00}  +  \cdots
\label{ai.pergt}
\end{equation}
where $\cdots$ in the RHS of Eq.~(\ref{ein.pergt}) and Eq.~(\ref{ai.pergt}) denote the additional terms in the RHS of Eq.~(\ref{ein.pertfg}) and Eq.~(\ref{ai.pertfg}) respectively. We see that Eq.~(\ref{ein.pergt})  and Eq.~(\ref{ai.pergt}) are simply of the form of Eq.~(\ref{scalar.fe}). Using Eq.~(\ref{ft.w}) we Fourier transform the independent perturbed fields $e_{ij}$ and $a_i$
\begin{align}
e_{ij}\left(t,\vec{x}\right) = \int\, d\omega\,e^{-i \omega t}\,\tilde{e}_{ij}\left(\omega,\vec{x}\right) \,,\qquad a_{i}\left(t,\vec{x}\right) = \int d\omega\, e^{-i \omega t}\, \tilde{a}_{i}\left(\omega,\vec{x}\right) \,,
\label{ff.ft}
\end{align}
Substituting Eq.~(\ref{ff.ft}) in Eq.~(\ref{ein.pertfg}) gives us
\begin{align}
 -T^{(P)}_{i j} &= -m\int \delta_{ki}\delta_{lj}\frac{\delta^4(x - r(\sigma))}{1 + 2\phi\left(\vec{r}\right)}\frac{dr^k}{d \sigma}\frac{dr^l}{d \sigma}\,d \sigma = \int d\omega\, e^{-i \omega t}\, \widetilde{\Box}\left(\left(1+2\phi\right) \tilde{e}_{ij}\right) \notag\\
& \quad - \int d\omega\, e^{-i \omega t}\, 4 \left[ \omega^2 \phi \tilde{e}_{ij} + i \omega \left(\phi_{,i} \tilde{e}_{j0} + \phi_{,j} \tilde{e}_{i0}\right) -\frac{1}{2}\left(\phi_{,ij} - \frac{1}{2}\phi_{,kk}\delta_{ij}\right) \left(\tilde{e}_{00}+ \tilde{e}_{ll}\right)\right] \notag\\
&\qquad + \frac{2 Q}{M}\int d\omega\, e^{-i \omega t} \left[i \omega \left(\tilde{a}_{i}\phi_{,j} + \tilde{a}_{j}\phi_{,i} - \delta_{ij} \tilde{a}_{l}\phi_{,l}\right) - 2 \tilde{a}_0\left(\phi_{,ij} - \frac{1}{2}\delta_{ij}\phi_{,kk}\right)\right]
\label{ft.grav}
\end{align}
and substituting Eq.~(\ref{ff.ft}) in Eq.~(\ref{ai.pertfg}) gives
\begin{align}
-\left(1 - \phi\right) 4 \pi J^{(P)}_i &= -q\int \delta_{ki}\frac{\delta^4(x - r(\sigma))}{1 + \phi\left(\vec{r}\right)}\frac{dr^k}{d \sigma}\,d \sigma = \int d\omega\, e^{-i \omega t}\, \widetilde{\Box}\left(\left(1+\phi\right)\tilde{a}_i\right) \notag\\ 
& \hspace{-1 em} - \int d\omega\, e^{-i \omega t} \left[ 4 \left(\omega^2 \tilde{a}_{i}\phi + i \omega  \tilde{a}_{0}\phi_{,i}\right) + \tilde{a}_i\phi_{,kk} - 2\tilde{a}_k\phi_{,ik} - \frac{16 \pi Q}{M}\tilde{e}_{0j}\phi_{,ij}\right]
\label{ft.em}
\end{align}
Comparing the first lines of Eq.~(\ref{ft.grav}) and Eq.~(\ref{ft.em}) with Eq.~(\ref{psi0.ftw}) allows us to determine the analog of the leading contribution $\tilde{\psi}^{(0)}_0$ in these equations, whose solution is given in Eq.~(\ref{psi00.solf}). Choosing $\tilde{\psi}^{(0)}_0(\omega\,,\vec{x}) =  \left(1 + 2 \phi \left(\vec{x}\right)\right) \tilde{e}_{ij}\left(\omega\,,\vec{x}\right)$ and $f(\sigma) = \delta_{ki}\delta_{lj}\frac{m}{1 + 2\phi\left(\vec{r}\right)}\frac{dr^k}{d \sigma}\frac{dr^l}{d \sigma}$, we find on comparing with Eq.~(\ref{psi0.ftw}) and Eq.~(\ref{psi00.solf}) the following leading order contribution $\tilde{e}^{(0)}_{ij}(\omega\,,\vec{x})$ in the solution of Eq.~(\ref{ein.pergt}) in frequency space
\begin{equation}
\tilde{e}^{(0)}_{ij}(\omega\,,\vec{x}) = \frac{m}{1 + 2 \phi(\vec{x})} \int \frac{e^{i \omega\left(r^0 + R_0\right)}}{4 \pi R_0} \frac{v_i v_j}{1+ 2 \phi(\vec{r})} \, \frac{dr^0}{d\sigma}dr^0 \,,
\label{lgrav.sol}
\end{equation}
where we changed the integration variable from $\sigma$ to $r^0$ in the final expression of the solution and have denoted $\frac{dr^k}{d r^0}$ as $v^k$.
Likewise, on choosing $\tilde{\psi}^{(0)}_0(\omega\,,\vec{x}) =  \left(1 + \phi \left(\vec{x}\right)\right) \tilde{a}_{i}\left(\omega\,,\vec{x}\right)$ and $f(\sigma) = \delta_{ki}\frac{q}{1 + \phi\left(\vec{r}\right)}\frac{dr^k}{d \sigma}$, we can compare with Eq.~(\ref{psi0.ftw}) and Eq.~(\ref{psi00.solf}) to find the following leading order contribution $\tilde{a}^{(0)}_{i}(\omega\,,\vec{x})$ in the solution of Eq.~(\ref{ai.pergt})
\begin{equation}
\tilde{a}^{(0)}_{i}(\omega\,,\vec{x}) = \frac{q}{1 + \phi(\vec{x})} \int \frac{e^{i \omega\left(r^0 + R_0\right)}}{4 \pi R_0} \frac{v_i}{1+ \phi(\vec{r})} \,dr^0 \,,
\label{lem.sol}
\end{equation}

Eq.~(\ref{lgrav.sol}) and Eq.~(\ref{lem.sol}) would be the complete solutions of Eq.~(\ref{ft.grav}) and Eq.~(\ref{ft.em}), respectively, in the absence of the second and third lines of Eq.~(\ref{ft.grav}) and the second line of Eq.~(\ref{ft.em}). The solution in the presence of these additional terms can be determined in the following way. First, since Eq.~(\ref{ft.grav}) and Eq.~(\ref{ft.em}) can be viewed as involving corrections for the equation for the scalar Green function, we can make use of Eq.~(\ref{gf.rn}) for all terms in Eq.~(\ref{lgrav.sol}) and Eq.~(\ref{lem.sol}) apart from those in the first line. Second, as these terms all involve $\phi$ or their derivatives, we can consider the gauge conditions in Eq.~(\ref{k.exp}) and Eq.~(\ref{l.exp}) to lowest order in $\phi$. The resulting equations from the gravitational gauge condition in Eq.~(\ref{k.exp}) are
\begin{align}
e_{ij,j} - e_{i0,0} = 0 \,, \qquad
e_{0i,i} - e_{00,0} =0 \,,
\label{dd.eq}
\end{align}
while the equation resulting from Eq.~(\ref{l.exp}) is
\begin{equation}
a_{0,0} - a_{i,i} = 0
\label{lor.eq}
\end{equation}
These are simply the de Donder and Lorentz gauges in flat spacetime. 
By Fourier transforming Eq.~(\ref{dd.eq}) and using Eq.~(\ref{lgrav.sol}), we derive the following lowest order in $\phi$ solutions
\begin{align}
\tilde{e}^{(0)}_{ij}(\omega\,,\vec{x}) &= m \int \frac{e^{i \omega\left(r^0 + R_0\right)}}{4 \pi R_0} v_i v_j \, \frac{dr^0}{d\sigma}dr^0 + \mathcal{O}(\phi)\,,\notag\\
\tilde{e}^{(0)}_{i0}(\omega\,,\vec{x}) &= -m \int \frac{e^{i \omega\left(r^0 + R_0\right)}}{4 \pi R_0} v_i  \, \frac{dr^0}{d\sigma}dr^0 + \mathcal{O}(\phi)\,,\notag\\
\tilde{e}^{(0)}_{00}(\omega\,,\vec{x}) &= m \int \frac{e^{i \omega\left(r^0 + R_0\right)}}{4 \pi R_0}  \, \frac{dr^0}{d\sigma}dr^0 + \mathcal{O}(\phi)\,.
\label{low.e}
\end{align}
Using Eq.~(\ref{lem.sol}) with the Fourier transform in Eq.~(\ref{lor.eq}) identifies the lowest order in $\phi$ solutions for electromagnetic perturbations
\begin{align}
\tilde{a}^{(0)}_{i}(\omega\,,\vec{x}) &= q \int \frac{e^{i \omega\left(r^0 + R_0\right)}}{4 \pi R_0} v_i \, dr^0 + \mathcal{O}(\phi)\,,\notag\\
\tilde{a}^{(0)}_{0}(\omega\,,\vec{x}) &= -q \int \frac{e^{i \omega\left(r^0 + R_0\right)}}{4 \pi R_0} dr^0 + \mathcal{O}(\phi)\,.
\label{low.a}
\end{align}
Hence the gauge conditions determine all the lowest order expressions. Substituting Eq.~(\ref{low.e}) and Eq.~(\ref{low.a}) in Eq.~(\ref{ft.grav}) and Eq.~(\ref{ft.em}), we then use the expressions in Eq.~(\ref{gf.rn}) and Eq.~(\ref{gf.rn2}) to determine the following solutions for $\tilde{e}_{ij}\left(\omega,\vec{x}\right)$ and $\tilde{a}_{i}\left(\omega,\vec{x}\right)$
\begin{align}
\tilde{e}_{ij}\left(\omega,\vec{x}\right) &= \frac{m}{1 + 2 \phi(\vec{x})} \int \frac{e^{i \omega\left(r^0 + R_0\right)}}{4 \pi R_0} \frac{v_i v_j}{1+ 2 \phi(\vec{r})} \, \frac{dr^0}{d\sigma}dr^0 \notag\\
&\quad - \int dr^0\, e^{i \omega r^0} \int d^3\vec{r}\,'\, \delta^{(3)} \left(\vec{r}\,' - \vec{r}\left(r^0\right)\right) \Bigg\{\frac{dr^0}{d\sigma} \frac{m}{\pi} \left[\omega^2 v^i v^j - i \omega \left(v^i\nabla_j + v^j\nabla_i\right) \phantom{\left(\frac{1}{2}\right)} \right. \notag\\
&\left. \qquad - \frac{\left(1+\vec{v}^2\right)}{2}\left(\nabla_i \nabla_j - \frac{1}{2}\delta_{ij}\nabla^2\right)\right] + \frac{q Q}{2 \pi M} \left[i\omega \left( v^i \nabla_j + v^j\nabla_j - \delta_{ij} v^k \nabla_k\right) \phantom{\left(\frac{1}{2}\right)}\right. \notag\\
& \left. \qquad \qquad \quad \, + 2 \left(\nabla_i \nabla_j - \frac{1}{2}\delta_{ij}\nabla^2\right)\right]\Bigg\} G\left(\omega,\vec{x},\vec{r}\,'\right)\,,
\label{hij.fin}
\end{align}
\begin{align}
\tilde{a}_{i}\left(\omega,\vec{x}\right) &= \frac{q}{1 + \phi(\vec{x})} \int \frac{e^{i \omega\left(r^0 + R_0\right)}}{4 \pi R_0} \frac{v_i }{1+ \phi(\vec{r})} \,dr^0 \notag\\
&\quad - \frac{q}{\pi} \int dr^0\, e^{i \omega r^0} \int d^3\vec{r}\,'\, \delta^{(3)} \left(\vec{r}\,' - \vec{r}\left(r^0\right)\right) \Bigg\{ \omega^2 v^i - i \omega \nabla_i  + \frac{dr^0}{d\sigma}\frac{4 Q m \pi}{M q} v^j\nabla_i\nabla_j \notag\\
& \qquad \qquad \qquad \qquad + \frac{1}{4} v^i\nabla^2 - \frac{1}{2} v^k \nabla_k \nabla_i \Bigg\} G\left(\omega,\vec{x},\vec{r}\,'\right)\,,
\label{ai.fin}
\end{align}
where $\nabla_i$ denotes $\frac{\partial}{\partial x^i} + \frac{\partial}{\partial r\,'^i}$ in Eq.~(\ref{hij.fin}) and Eq.~(\ref{ai.fin}). In considering $Q=0$, Eq.~(\ref{hij.fin}) agrees with \cite{Peters:1970mx} and Eq.~(\ref{ai.fin}) agrees with \cite{Peters:1973ah}, after taking into account the change in notation and metric signature. The terms involving $Q$ are hence additional contributions due to the RN spacetime. In the following section, we consider the explicit evaluation of the integrals involved in the radiative solutions Eq.~(\ref{hij.fin}) and Eq.~(\ref{ai.fin}) in the limit where $\omega \to 0$. This will enable us to verify that the above results satisfy the soft theorem in the presence of gravitational and electromagnetic interactions and provide the known tree level soft factors in this case.

\section{Soft factors and waveforms from classical radiation} \label{soft}

In \cite{Laddha:2018rle}, by considering the classical limit of multiple soft theorems in four and higher dimensional spacetimes, the soft factor up to subleading order was related with the power spectrum of low frequency radiation emitted in classical scattering processes. This relation was further used in \cite{Laddha:2018myi} to unambiguously identify the soft factor in four spacetime dimensions from scattering processes involving a heavy center and a probe particle. The soft factor in this case develop certain logarithmic contributions arising from the long-range gravitational and electromagnetic forces experienced by the soft particles. In particular, the result of \cite{Peters:1970mx} was shown to satisfy the predicted soft factor and also helped identify an overall phase of the soft radiation. This phase can be understood as arising from the gravitational drag experienced by the soft particle due to the presence of the large scatterer. While the overall phase does not affect the flux of soft particles, it does have implications on the late time gravitational waveform \cite{Laddha:2018vbn}. The soft factor determined from classical scattering processes was shown to be consistent with quantum results \cite{Sahoo:2018lxl}.

In the following subsections, we will first review the results of \cite{Laddha:2018rle,Laddha:2018myi,Sahoo:2018lxl}. Following these references, we will then evaluate the integrals involved in Eq.~(\ref{hij.fin}) and Eq.~(\ref{ai.fin}) in the limit where $\vec{x} \gg \vec{r}\left(\sigma\right)$ and when $\omega \to 0$. The resulting expressions will be shown to provide the correct soft factor involved in the soft photon and soft graviton theorem in the presence of gravitational and electromagnetic interactions. Following the analysis of \cite{Laddha:2018vbn}, we investigate the implications of the soft factor results on the memory and tail effect of late-time gravitational and electromagnetic waveforms in the last subsection.

\subsection{Soft factors and evaluation of integrals in four-dimensional spacetimes}

The general relation between the tree level or classical soft factor and the classical electromagnetic and gravitational radiation in $D$ spacetime dimensions was provided in~\cite{Laddha:2018rle}. Given the trace reversed metric perturbation $\tilde{e}_{\alpha \beta}(\omega\,, \vec{x})$ and electromagnetic field $\tilde{a}_a(\omega\,, \vec{x})$ in frequency space, we have
\begin{align}
\epsilon^{\alpha \beta}\tilde{e}_{\alpha \beta}(\omega\,, \vec{x}) &= \mathcal{N}' S_{\text{gr}}\left(\epsilon\,,k\right) \label{soft.gr}\\
\epsilon^{\alpha}\tilde{a}_{\alpha}(\omega\,, \vec{x}) &= \mathcal{N}' S_{\text{em}}\left(\epsilon\,,k\right) \,, \label{soft.em} \\
\mathcal{N}' = \frac{1}{2 \omega}e^{i \omega R} \left( \frac{\omega}{2 \pi i R}\right)^\frac{D-2}{2}\,, \quad R &= \vert \vec{x}\vert\,, \quad  \hat{n} = \frac{\vec{x}}{R}\,, \quad k = - \omega \left(1,\hat{n}\right)\,.  \label{soft.norm}
\end{align}
In Eq.~(\ref{soft.gr}) $\epsilon^{\alpha \beta}$ denotes an arbitrary rank two polarization tensor and $k$ denotes the momentum of the soft graviton, while in Eq.~(\ref{soft.em}) $\epsilon^{\alpha}$ denotes an arbitrary polarization vector and $k$ denotes the momentum of the soft photon. $S_{\text{em}}$ and $S_{\text{gr}}$ are the soft factors in the classical limit of the soft graviton theorem and soft photon theorem with the expressions
\begin{align}
S_{\text{gr}} =S^{(0)}_{\text{gr}} + S^{(1)}_{\text{gr}} \,, \qquad & S_{\text{em}} =S^{(0)}_{\text{em}} + S^{(1)}_{\text{em}} \notag\\
S^{(0)}_{\text{gr}} = \sum_{a=1}^n \frac{\epsilon_{\mu \nu} p^{\mu}_{(a)}p^{\nu}_{(a)}}{p_{(a)}.k} \,, & \quad S^{(1)}_{\text{gr}} = i \sum_{a=1}^n \frac{\epsilon_{\mu \nu} p^{\mu}_{(a)}k_\rho}{p_{(a)}.k} {\bf{J}}^{\rho \nu}_{(a)} \label{gr.lsl}\\
S^{(0)}_{\text{em}} = \sum_{a=1}^n q_{(a)}\frac{\epsilon. p_{(a)}}{p_{(a)}.k} \,, & \quad S^{(1)}_{\text{em}} = i \sum_{a=1}^n q_{(a)} \frac{\epsilon_{\nu} k_\rho}{p_{(a)}.k} {\bf{J}}^{\rho \nu}_{(a)} + \text{non-universal} \label{em.lsl}
\end{align}
$S^{(0)}$ and $S^{(1)}$ denote the leading and subleading contributions, respectively, in the soft factors. In the case of the subleading soft photon factor, there do exist non-universal contributions and the expression in Eq.~(\ref{em.lsl}) denotes only the universal piece. The sum in Eq.~(\ref{gr.lsl}) and Eq.~(\ref{em.lsl}) run over all incoming and outgoing particles, where $q_{(a)}$ $p_{(a)}$ and ${\bf{J}}_{(a)}$ denote the charges, momenta and angular momenta, counted with a positive sign for incoming particles and a negative sign for outgoing particles.

In~\cite{Laddha:2018myi}, the expressions for $S_{\text{gr}}$ and $S_{\text{em}}$ were determined for a system comprising of a probe particle of mass $m$ and charge $q$ scattering off a heavy central object of either mass $M$ or charge $Q$ in four spacetime dimensions. The result for $S_{\text{gr}}$ in the case of only gravitational interactions was also shown to agree with~\cite{Peters:1970mx}  after evaluating the integrals involved in the classical solutions. We will now review this procedure and apply it subsequently to the gravitational and electromagnetic radiative solutions on the RN spacetime. 

The expression for $S_{\text{gr}}$ in Eq.~(\ref{soft.gr}) is invariant under $\epsilon^{\mu\nu} \to \epsilon^{\mu \nu} + \xi^{\mu}k^{\nu} + \xi^{\nu}k^{\mu}$, where $\xi^{\alpha}$ is arbitrary. This implies the constraint $k^{\mu} \tilde{e}_{\mu\nu} = 0$ for the radiative components of the gravitational field. We can hence determine the $\tilde{e}_{0 \nu}$ components from $\tilde{e}_{ij}$ and set $\epsilon^{0\nu} = 0$. Similarly, $S_{\text{em}}$ in Eq.~(\ref{soft.em}) is invariant under $\epsilon^{\mu} \to \epsilon^{\mu} + k^{\mu}$. The constraint $k^{\mu}\tilde{a}_{\mu} = 0$ on the radiative components imply that $\tilde{a}_{0}$ can be determined from $\tilde{a}_i$ and that we can set $\epsilon^0 = 0$. We thus have
\begin{equation}
\epsilon^{0 \nu} = 0 \,, \qquad \qquad \epsilon^0 = 0
\label{pol.g}
\end{equation}
In addition, the expressions for $\tilde{e}_{\mu\nu}$ and $\tilde{a}_{\mu}$ are known only up to a choice in gauge. Denoting the arbitrary gauge parameters by $\xi_{\mu}$ and $\xi$ (for gravity) and $\lambda$ (for electromagnetism) , we have the following gauge transformations
\begin{equation}
\delta \tilde{e}_{\mu \nu} =  k_{\mu}\xi_{\nu} + k_{\nu}\xi_{\mu} - \xi. k \eta_{\mu\nu} \,, \qquad \delta \tilde{a}_{\mu} = \lambda k_{\mu} \,. 
\label{emgr.gtf}
\end{equation}
Using Eq.~(\ref{emgr.gtf}) in Eq.~(\ref{soft.gr}) and Eq.~(\ref{soft.em}), we then have the relations
\begin{equation}
k_{\mu}\epsilon^{\mu\nu} - \frac{1}{2}k^{\nu}\epsilon^{\mu}_{\mu} = 0 \,, \qquad k_{\mu}\epsilon^{\mu} = 0 \,.
\label{rad.gc}
\end{equation}
Eq.~(\ref{rad.gc}) and Eq.~(\ref{pol.g}) then provide the following conditions on the polarization tensor in Eq.~(\ref{soft.gr}) and the polarization vector in Eq.~(\ref{soft.em})
\begin{equation}
\epsilon^{0 \nu} = 0 \,,\quad  k_i \epsilon^{ij} = 0 \,, \quad \epsilon^i_i =0 \,; \qquad \qquad \qquad \epsilon^0 = 0 \,,\quad  k_i \epsilon^i =0\,.
\label{pol.cond}
\end{equation}

To determine $S_{\text{gr}}$ and $S_{\text{em}}$, we also need to determine the expressions for the momenta and angular momenta. Due to the presence of the long range interaction of electromagnetic and gravitational forces, the particle trajectories of the probe before ($r_{(1)}$) and after ($r_{(2)}$) the scattering can be taken to have the form
\begin{equation}
r^0_{(1)} = t = r^0_{(2)}\,, \qquad \vec{r}_{(1)} = \vec{\beta}_{-} t + \vec{c}_{-} - C_{-} \vec{\beta}_{-} \ln \vert t \vert\,, \qquad \vec{r}_{(2)} = \vec{\beta}_{+} t + \vec{c}_{+} - C_{+} \vec{\beta}_{+} \ln \vert t \vert\,,
\label{p.traj}
\end{equation}
where the $\ln \vert t \vert$ terms are the contributions to the long range forces. The momenta of the probe particle before and after scattering will be denoted by $p_{(1)}$ and $p_{(2)}$ respectively, while the angular momenta of the probe before and after scattering will be denoted by ${\bf{j}}^{\mu\nu}_{(1)}$ and ${\bf{j}}^{\mu\nu}_{(2)}$ respectively. Using Eq.~(\ref{p.traj}) and retaining all $\ln \vert t \vert$ terms, the following asymptotic expressions for the momenta and angular momenta can be determined
\begin{align}
p_{(1)} = \frac{m}{\sqrt{1 - \vec{\beta}_-^2}}\left(1\,, \vec{\beta}_-\right) \,, \qquad & p_{(2)} = -\frac{m}{\sqrt{1 - \vec{\beta}_+^2}}\left(1\,, \vec{\beta}_+\right) \,, \notag\\
{\bf{j}}^{ij}_{(1)} = r^i_{(1)} p^j_{(1)} - r^j_{(1)} p^i_{(1)} &= \frac{m}{\sqrt{1 - \vec{\beta}_-^2}} \left( c_-^i \beta_-^j -  c_-^j \beta_-^i\right) \,,  \notag\\
{\bf{j}}^{0i}_{(1)} = r^0_{(1)} p^i_{(1)} - r^i_{(1)} p^0_{(1)} &= - \frac{m}{\sqrt{1 - \vec{\beta}_-^2}} \left( c_-^i - C_- \beta^i_- \ln \vert t \vert\right)\,, \notag\\
{\bf{j}}^{ij}_{(2)} = r^i_{(2)} p^j_{(2)} - r^j_{(2)} p^i_{(2)} &= -\frac{m}{\sqrt{1 - \vec{\beta}_+^2}} \left( c_+^i \beta_+^j -  c_+^j \beta_+^i\right) \,,  \notag\\
{\bf{j}}^{0i}_{(2)} = r^0_{(2)} p^i_{(2)} - r^i_{(2)} p^0_{(2)} &= \frac{m}{\sqrt{1 - \vec{\beta}_+^2}} \left( c_+^i - C_+ \beta^i_+ \ln \vert t \vert\right)\,.
\label{mam.pp}
\end{align}
Substituting Eq.~(\ref{pol.cond}), Eq.~(\ref{mam.pp}) and the expression for $k$ from Eq.~(\ref{soft.norm}) in Eq.~(\ref{soft.gr}) and Eq.~(\ref{soft.em}), and replacing $\ln \vert t \vert$ with $\ln \omega^{-1}$, gives the expressions for the soft factors in the probe-scatterer approximation up to overall phases
\begin{align}
S_{\text{em}} &= - \frac{q}{\omega}\left[\frac{\vec{\epsilon}. \vec{\beta}_+}{1 - \hat{n}.\vec{\beta}_+} - \frac{\vec{\epsilon}. \vec{\beta}_-}{1 - \hat{n}.\vec{\beta}_-}\right] - i q \ln \omega^{-1} \left[C_+ \frac{\vec{\epsilon}.\vec{\beta}_+}{1 - \hat{n}.\vec{\beta}_+} - C_- \frac{\vec{\epsilon}.\vec{\beta}_-}{1 - \hat{n}.\vec{\beta}_-}\right] + \text{finite} \label{sem.pp}\\
S_{\text{gr}} &= - \frac{m}{\omega} \epsilon^{ij}\left[\frac{1}{1 - \hat{n}.\vec{\beta}_+}\frac{1}{\sqrt{1 - \vec{\beta}^2_+}}\beta_{+i} \beta_{+j} - \frac{1}{1 - \hat{n}.\vec{\beta}_-}\frac{1}{\sqrt{1 - \vec{\beta}^2_-}}\beta_{-i} \beta_{-j}\right] \notag\\
&-i m \ln \omega^{-1} \epsilon^{ij} \left[C_+\frac{1}{1 - \hat{n}.\vec{\beta}_+}\frac{1}{\sqrt{1 - \vec{\beta}^2_+}}\beta_{+i} \beta_{+j} - C_-\frac{1}{1 - \hat{n}.\vec{\beta}_-}\frac{1}{\sqrt{1 - \vec{\beta}^2_-}}\beta_{-i} \beta_{-j}\right] + \text{finite} \label{sgr.pp}
\end{align}
The substitution of $\ln \vert t \vert$ with $\ln \omega^{-1}$ was confirmed by comparison with classical scattering examples. This included the solution for gravitational radiation given in~\cite{Peters:1970mx}, which further provided evidence for an overall phase present in the soft factor arising from the backscattering of soft gravitons due to the potential of the central scatterer. The presence of $\ln \omega^{-1}$ in the quantum subleading soft factors and a first principle account for the overall phase in four dimensions were provided in~\cite{Sahoo:2018lxl}. The classical result therein involves a covariant generalization beyond the probe-scatterer approximation in the presence of both gravitational and electromagnetic interactions. The phase term for the low frequency gravitational waves was argued to arise as a consequence of logarithmic corrections in the trajectory of the soft particle. Specifically, the soft particle trajectory is
\begin{equation}
x^{\mu}\left(\tau\right) = n^{\mu}\left(\tau\right) + m^{\mu} \ln \vert \tau \vert \,,
\end{equation}
where $\tau$ its affine parameter associated with the trajectory, $n = \left(1,\hat{n}\right)$ is a null vector along the asymptotic direction of motion of the soft particle and $m^{\mu}$ is a four vector, which in the case of a heavy central scatterer of mass $M$ is given by
\begin{equation}
m^{\mu} = \frac{M}{4 \pi}\left(1, \vec{0}\right)
\end{equation}
By using the equation of motion of the soft particle, the analysis of~\cite{Sahoo:2018lxl} identifies the overall phase 
\begin{equation}
\exp [i \theta] = \exp \left[i k.m \ln \left(R \omega\right)\right]\,,
\label{ss.phase}
\end{equation}
where $R$ is the distance of the soft particle from the scattering center. This is the common overall phase which multiplies the soft photon and graviton factors, which in the probe-scatterer approximation becomes $\exp\left[ i \frac{M}{4 \pi} \omega \ln \left(R \omega\right)\right]$. By expanding the exponent and keeping terms of order $\omega \ln \left(R \omega\right)$, one finds a phase correction at the subleading soft photon and graviton factors. The corrections to Eq.~(\ref{sem.pp}) and Eq.~(\ref{sgr.pp}) are
\begin{align}
\Delta S_{\text{em}} &= -i q \frac{M}{4 \pi} \ln \left(R \omega\right) \left[\frac{\vec{\epsilon}. \vec{\beta}_+}{1 - \hat{n}.\vec{\beta}_+} - \frac{\vec{\epsilon}. \vec{\beta}_-}{1 - \hat{n}.\vec{\beta}_-}\right] \label{semc.pp}\\
\Delta S_{\text{gr}} &= -i m \frac{M}{4 \pi} \ln \left(R \omega\right) \epsilon^{ij}\left[\frac{1}{1 - \hat{n}.\vec{\beta}_+}\frac{1}{\sqrt{1 - \vec{\beta}^2_+}}\beta_{+i} \beta_{+j} - \frac{1}{1 - \hat{n}.\vec{\beta}_-}\frac{1}{\sqrt{1 - \vec{\beta}^2_-}}\beta_{-i} \beta_{-j}\right]
\label{sgrc.pp}
\end{align}
The expressions for $S_{\text{em}}+ \Delta S_{\text{em}}$ and  $S_{\text{gr}}+ \Delta S_{\text{gr}}$, using Eq.~(\ref{sem.pp}),Eq.~(\ref{sgr.pp}),Eq.~(\ref{sgrc.pp}) and Eq.~(\ref{semc.pp}), agree with the soft factor expressions of~\cite{Sahoo:2018lxl} in the probe-scatterer approximation.

In considering the classical expressions for the radiative fields, we will need to evaluate certain integrals of the type
\begin{equation}
I = \int dt e ^{i \omega g(t)} F(t) + \text{boundary terms} \,,
\label{soft.int}
\end{equation}
where $g(t) \to a_{\pm} t + b_{\pm} \ln \vert t \vert$ and $F(t)$ is either a constant or falls off as a negative power of t as $t \to  \infty$. 
 If $F(t) \sim \vert t \vert^{- \alpha}$ with $-1 < \alpha \le 0$ ($\alpha = 0$ is the case where $F(t)$ is a constant)
then the integral in Eq.~(\ref{soft.int}) has to be defined by first performing an integration by parts using the identity $e ^{i \omega g(t)} = \frac{1}{i \omega g'(t)} \frac{d}{dt} e ^{i \omega g(t)}$, where $g'(t) = \frac{d}{dt}g(t)$. This gives
\begin{equation}
I = \int dt \frac{1}{i \omega g'(t)} \frac{d}{dt} e ^{i \omega g(t)} F(t) + \text{boundary terms} = - \frac{1}{i \omega} \int dt e ^{i \omega g(t)} \frac{d}{dt} \left(\frac{F(t)}{i \omega g'(t)}\right)
\label{soft.int2}
\end{equation}
In evaluating the integrals, we will make specific use of the following five integral relations derived in~\cite{Laddha:2018myi}
\begin{align}
I_1 &= \frac{1}{\omega} \int \limits_{-\infty}^{+\infty} dt e^{-i \omega g(t)} \frac{d}{dt}f(t) = \frac{1}{\omega} \left(f_{+} - f_{-}\right) + i \left(a_{+}k_{+} - a_{-}k_{-}\right) \text{ln} \omega^{-1} + \text{finite}\notag\\
I_2 &= \int \limits_{-\infty}^{+\infty} dt e^{-i \omega g(t)} \frac{d}{dt}\left[f(t)\left(\text{ln}\frac{h(t)}{R} + \int\limits_{h(t)}^{+\infty}du \frac{e^{i \omega u}}{u}\right)\right] = -\left(f_{+} - f_{-}\right) \text{ln}\left(R\omega\right)+ \text{finite}\notag\\
I_3 &= \int \limits_{-\infty}^{+\infty} dt \frac{f(t)}{r(t)}\left(e^{-i \omega g(t)} - e^{-i\omega h(t)}\right) = \text{finite} \notag\\
I_4 &= \frac{1}{\omega} \int \limits_{-\infty}^{+\infty} dt \frac{f(t)}{r(t)^2}\left(e^{-i \omega g(t)} - e^{-i \omega h(t)}\right) = -i\left(\frac{f_+}{c_+^2} \left(a_{+} - p_{+}\right) - \frac{f_-}{c_-^2}\left(a_{-}- p_{-}\right)\right) \text{ln} \omega^{-1} + \text{finite}\notag\\
I_5 &= \int \limits_{-\infty}^{+\infty} dt \frac{f(t)}{r(t)}e^{-i \omega g(t)} = \left(\frac{f_{+}}{c_+} - \frac{f_{-}}{c_-}\right)\text{ln} \omega^{-1} + \text{finite} \,,
\label{int.exp}
\end{align}
where
\begin{align}
f(t) &\to f_{\pm} + \frac{k_{\pm}}{t} \,, \qquad g(t) \to a_{\pm} t + b_{\pm} \ln \vert t \vert \notag\\
h(t) &\to p_{\pm} t + q_{\pm} \ln \vert t \vert \,, \quad r(t) \to c_{\pm} t + d_{\pm} \ln \vert t \vert \,, \qquad \text{as}\, t\to \infty\,,
\end{align}
In Eq.~(\ref{int.exp}) `$R$' is an arbitrary constant and `finite' represent terms which are finite in the $\omega \to 0$ limit. The integral relations in Eq.~(\ref{int.exp}) were used to demonstrate that the result of~\cite{Peters:1970mx} agrees with the graviton soft factor. In the following subsections, we will demonstrate that the photon and graviton soft factors are also satisfied by the probe particle scattering results on the RN spacetime.

\subsection{Electromagnetic soft factors and phases on the RN spacetime}\label{electrosoft}

In this subsection we will evaluate Eq.~(\ref{ai.fin}) in order to determine the soft factor expression. Using Eq.~(\ref{soft.em}), the normalization in $D=4$ dimensions from Eq.~(\ref{soft.norm}) and the polarization conditions Eq.~(\ref{pol.cond}), we have the following four dimensional expression for the electromagnetic soft factor
\begin{equation}
S_{\text{em}} = i \frac{4 \pi R}{e^{i \omega R}} \epsilon^{i}\tilde{a}_{i} (\omega, \vec{x}) \,. \label{sfact4.em}
\end{equation}
The integrals in Eq.~(\ref{ai.fin}) will be evaluated in the limit where the electromagnetic waves are far separated from the probe particle, i.e. $\vec{x} >> \vec{r}\left(\sigma\right)$. In this limit, the expression of $G_M\left(\omega,\vec{x},\vec{r}\right)$ in Eq.~(\ref{gfn}) simplifies to
\begin{align}
\lim_{\vec{x}\gg\vec{r}} G_M\left(\omega,\vec{x},\vec{r}\right) \rightarrow \frac{i M}{16 \pi \omega} \left[\ln\left(\frac{\vert\vec{r}\vert+ \hat{n}.\vec{r}}{R}\right) + \int_{|\vec{r}|+\hat{n}.\vec{r}}^\infty\frac{du}{u} e^{i\omega u}\right] \frac{e^{i\omega (R - \hat{n}.\vec{r})}}{R}\,:=\tilde{G}_M\left(\omega,\vec{x},\vec{r}\right),
\label{gfn.lim}
\end{align}
Setting $r^0 = t$ and using Eq.~(\ref{gfn.lim}) in Eq.~(\ref{ai.fin}), we find
that the solutions for electromagnetic perturbations in Eq.~(\ref{ai.fin}) take the form
\begin{align}
\tilde{a}_i(\omega, \vec{x})&=\tilde{a}_i^{(1)}(\omega, \vec{x})+\tilde{a}_i^{(2)}(\omega, \vec{x})+\tilde{a}_i^{(3)}(\omega, \vec{x})+\tilde{a}_i^{(4)}(\omega, \vec{x})+\tilde{a}_i^{(5)}(\omega, \vec{x}) + \tilde{a}_i^{(6)}(\omega, \vec{x}) \,,
\end{align}
where
\begin{align}
\tilde{a}_i^{(1)}(\omega\,,\vec{x})
&=\frac{q}{4\pi}\frac{e^{i\omega R}}{R}\int dt\, \frac{e^{i\omega(t-\hat{n}.\vec{r})}}{1+\phi(\vec{r}(t))}\,v_i \label{ai.1}
\end{align}
\begin{align}
\tilde{a}_i^{(2)}(\omega\,,\vec{x}) &= \frac{q}{2\pi}\int dt\, e^{i\omega t}\,v_k\nabla_k\nabla_i\tilde{G}_M\left(\omega,\vec{x},\vec{r}\right)\,, \label{ai.2}
\end{align}
\begin{align}
\tilde{a}_i^{(3)}(\omega\,,\vec{x}) &= \frac{i\omega q}{\pi}\int dt\, e^{i\omega t} \,\nabla_i\tilde{G}_M\left(\omega,\vec{x},\vec{r}\right)\,,\label{ai.3}
\end{align}
\begin{align}
\tilde{a}_i^{(4)}(\omega\,,\vec{x}) &=-\frac{iMq\omega}{16\pi^2}\frac{e^{i\omega R}}{R}\int dt v_i\bigg\{
\ln \frac{|\vec{r}\,'|+\hat{n}.\vec{r}\,'}{ R} \, e^{i\omega (t - \hat{n}.\vec{r}\,')} + \int_{|\vec{r}\,'|+\hat{n}.\vec{r}\,'}^\infty \frac{du}{u} e^{i\omega (t - \hat{n}.\vec{r}\,'+u)}\bigg\}\label{ai.4}
\end{align}
\begin{align}
\tilde{a}_i^{(5)}(\omega\,,\vec{x}) &=-\frac{q}{4\pi}\int dt\, {e^{i\omega t}}\, v_i\nabla_k\nabla_k\tilde{G}_M\left(\omega,\vec{x},\vec{r}\right)\label{ai.5}
\end{align}
\begin{align}
\tilde{a}_i^{(6)}(\omega\,,\vec{x}) &= \frac{4 Q m}{M} \int dt\, \frac{dt}{d\sigma}\, e^{i\omega t} \,v_k \nabla_k\nabla_i\tilde{G}_M\left(\omega,\vec{x},\vec{r}\right), \label{ai.6}
\end{align}
where $\nabla_i= \frac{\partial}{\partial r^i}+\frac{\partial}{\partial x^i}$. Apart from $\tilde{a}_i^{(6)}(\omega, \vec{x})$ term which involves $Q$, the terms $\tilde{a}_i^{(1)}(\omega, \vec{x})\,, \cdots \tilde{a}_i^{(5)}(\omega, \vec{x})$ are the same as those which arise for a probe charge on the Schwarzschild spacetime~\cite{Peters:1973ah}, after accounting for the difference in notation and metric signature.

As in Eq.~(\ref{p.traj}), we assume the following asymptotic expressions for the particle trajectory and its velocity as $t \to \pm \infty$
\begin{align}
\vec{r} = \vec{\beta}_{\pm} t + \vec{c}_{\pm} - C_{\pm} \vec{\beta}_{\pm} \ln \vert t \vert \,, \qquad \vec{v} = \frac{d\vec{r}}{dt} = \vec{\beta}_{\pm} \left(1 - \frac{C_{\pm}}{t}\right) \label{as.rv}
\end{align}

We will first evaluate $\tilde{a}^{(1)}_{i}(\omega, \vec{x})$. Since the integrand involves a constant piece, we need to perform an integration by parts
and Eq.~(\ref{ai.1}) becomes
\begin{equation} 
\tilde{a}_i^{(1)}(\omega\,,\vec{x})
= - \frac{q}{4\pi}\frac{e^{i\omega R}}{R}\frac{1}{i \omega}\int e^{i\omega(t-\hat{n}.\vec{r})} \frac{d}{dt}\left[\frac{1}{1+\phi(\vec{r})}\frac{1}{1 - \hat{n}.\vec{v}} v_i\right] dt \label{ai1.ibp}
\end{equation}

From Eq.~(\ref{as.rv}), we have 
\begin{equation}
\left(1 - \hat{n}.\vec{v}\right)^{-1} = \left(1 - \hat{n}.\vec{\beta}_{\pm}\right)^{-1} \left[1 - \frac{C_{\pm}}{t} \frac{\hat{n}.\vec{\beta}_{\pm}}{1 - \hat{n}.\vec{\beta}_{\pm}}\right] + \mathcal{O}\left(t^{-2}\right) \label{as.nv}
\end{equation}
and
\begin{equation}
\phi(\vec{r}(t)) = -\frac{M}{8 \pi \vert \vec{r}(t) \vert} = \mp \frac{M}{8 \pi \vert \vec{\beta}_{\pm} \vert t}   + \mathcal{O}\left(t^{-2}\right)\,, \label{as.pot}
\end{equation}

where the $\mp$ sign in the last equality follows from the sign convention used for incoming and outgoing states.
Using Eq.~(\ref{as.rv}), Eq.~(\ref{as.nv}) and Eq.~(\ref{as.pot}) in Eq.~(\ref{ai1.ibp}), we find the expression
\begin{align}
\tilde{a}_i^{(1)}(\omega\,,\vec{x}) &= \frac{i q e^{i\omega R}}{4\pi R}\frac{1}{\omega} \int e^{- i\omega\left(\left(\hat{n}.\vec{\beta}_{\pm} - 1\right)t - C_{\pm}\hat{n}.\vec{\beta}_{\pm}\ln \vert t\vert\right)} \frac{d}{dt}\left[\frac{\beta_{\pm i}}{1 - \hat{n}.\vec{\beta}_{\pm}}\left(1 - \frac{1}{t}\left(\frac{C_{\pm}}{1 - \hat{n}.\vec{\beta}_{\pm}} \mp \frac{M}{8 \pi \vert \vec{\beta}_{\pm}\vert}\right)\right)\right] dt \notag\\
&\qquad \qquad + \mathcal{O}\left(t^{-3}\right) \label{ai1.fin}
\end{align}
This integral is of the $I_1$ type in Eq.~(\ref{int.exp}) with the identifications
\begin{align}
f_{\pm} &= \frac{i q e^{i\omega R}}{4\pi R}\frac{\beta_{\pm i}}{1 - \hat{n}.\vec{\beta}_{\pm}} \,, \notag\\
k_{\pm} &= \frac{i q e^{i\omega R}}{4\pi R}\frac{\beta_{\pm i}}{1 - \hat{n}.\vec{\beta}_{\pm}}\left(\frac{C_{\pm}}{\hat{n}.\vec{\beta}_{\pm}-1} \pm \frac{M}{8 \pi \vert \vec{\beta}_{\pm}\vert}\right)\,,\notag\\
a_{\pm} & = \hat{n}.\vec{\beta}_{\pm} - 1 \,, \qquad b_{\pm} =  - C_{\pm}\hat{n}.\vec{\beta}_{\pm}
\end{align}
Hence we get the following contribution from Eq.~(\ref{ai1.fin})
\begin{align}
\tilde{a}_i^{(1)}(\omega\,,\vec{x}) &= i \frac{q}{\omega} \frac{e^{i\omega R}}{4 \pi R} \left(\frac{1}{1 - \hat{n}.\vec{\beta}_+}\beta_{+i} - \frac{1}{1 - \hat{n}.\vec{\beta}_-}\beta^i_-\right)\notag\\
& \qquad  - q \ln \omega^{-1} \frac{e^{i\omega R}}{4 \pi R}\left(\frac{C_+}{1 - \hat{n}.\vec{\beta}_+}\beta_{+i} - \frac{C_-}{1 - \hat{n}.\vec{\beta}_-}\beta_{-i} - \frac{M}{8 \pi \vert \vec{\beta}_+\vert}\beta_{+i} - \frac{M}{8 \pi \vert \vec{\beta}_-\vert}\beta_{-i}\right)
\label{ai1.exp}
\end{align}

Using Eq.~(\ref{sfact4.em}), we determine the following contribution from Eq.~(\ref{ai1.exp}) to the photon soft factor
\begin{align}
i \frac{4 \pi R}{e^{i \omega R}} \epsilon^{i}\tilde{a}_i^{(1)}(\omega, \vec{x}) &= - \frac{q}{\omega} \left(\frac{\vec{\epsilon}. \vec{\beta}_+}{1 - \hat{n}.\vec{\beta}_+} - \frac{\vec{\epsilon}. \vec{\beta}_-}{1 - \hat{n}.\vec{\beta}_-}\right)\notag\\
& \qquad  - i q \ln \omega^{-1} \left(\frac{C_+}{1 - \hat{n}.\vec{\beta}_+}\vec{\epsilon}. \vec{\beta}_+ - \frac{C_-}{1 - \hat{n}.\vec{\beta}_-}\vec{\epsilon}. \vec{\beta}_- - \frac{M}{8 \pi \vert \vec{\beta}_+\vert}\vec{\epsilon}. \vec{\beta}_+ - \frac{M}{8 \pi \vert \vec{\beta}_-\vert}\vec{\epsilon}. \vec{\beta}_-\right)
\label{ai1.soft}
\end{align}
From Eq.~(\ref{ai1.soft}), we see that $\tilde{a}_i^{(1)}(\omega, \vec{x})$ provides the leading term in the soft factor expression and a subleading contribution. 
As in the case of $\tilde{a}_i^{(1)}(\omega, \vec{x})$ in Eq.~(\ref{ai.1}), the integral for $\tilde{a}_i^{(4)}(\omega\,,\vec{x})$ in Eq.~(\ref{ai.4}) also requires an integration by parts, following which we have
\begin{align}
\tilde{a}_i^{(4)}(\omega, \vec{x}) = \frac{M q e^{i\omega R}}{16\pi^2 R}\int dt e^{i\omega(t-\hat{n}.\vec{r})} \frac{d}{dt}\left[ \frac{v_i}{1 - \hat{n}.\vec{v}} \bigg\{
\ln \frac{|\vec{r}|+\hat{n}.\vec{r}}{ R} + \int_{|\vec{r}|+\hat{n}.\vec{r}}^\infty \frac{du}{u} e^{i\omega u}\bigg\}\right]
\label{ai4.ibp}
\end{align}
Substituting Eq.~(\ref{as.rv}) in  Eq.~(\ref{ai4.ibp}), the integral takes the form of $I_2$ in Eq.~(\ref{int.exp}). Hence $\tilde{a}_i^{(4)}(\omega, \vec{x})$ provides the expression
\begin{equation}
\tilde{a}_i^{(4)}(\omega\,,\vec{x}) = - M q \frac{e^{i \omega R}}{16 \pi^2 R} \ln\left(\omega R\right) \left(\frac{1}{1-\hat{n}.\vec{\beta}_+}\beta_{+i} - \frac{1}{1-\hat{n}.\vec{\beta}_-}\beta_{-i}\right) + \text{finite}
\label{ai4.exp}
\end{equation}
Using Eq.~(\ref{sfact4.em}) with Eq.~(\ref{ai4.exp}), we find the following soft factor contribution
\begin{equation}
i \frac{4 \pi R}{e^{i \omega R}} \epsilon^{i}\tilde{a}_i^{(4)}(\omega\,,\vec{x})= - i q \frac{M}{4 \pi} \ln\left(\omega R\right) \left(\frac{\vec{\epsilon}. \vec{\beta}_+}{1-\hat{n}.\vec{\beta}_+} - \frac{\vec{\epsilon}. \vec{\beta}_-}{1-\hat{n}.\vec{\beta}_-}\right)
\label{ai4.soft}
\end{equation}

Eq.~(\ref{ai4.soft}) agrees with Eq.~(\ref{semc.pp}). Hence $\tilde{a}_i^{(4)}(\omega\,,\vec{x})$ provides a pure phase contribution to the soft factor expression.

We will next consider $\tilde{a}_i^{(3)}(\omega, \vec{x})$. For this term we require
\begin{align}
\nabla_i\tilde{G}_M= \frac{iM}{16\pi\omega}\frac{1}{|\vec{r}\,|+\hat{n}.\vec{r}\,}\left(\frac{r_i}{\vert \vec{r}\vert} + \hat{n}_i\right)\frac{1}{R}\bigg\{ e^{i \omega \left( R- \hat{n}\vec{r}\right)} -   e^{i \omega \left(R + \vert\vec{r}\vert\right)}\bigg\}
\label{as.der}
\end{align}
All terms involving $\hat{n}_i$ in this expression do not contribute to the soft factor since $\epsilon^{i}\hat{n}_i = \epsilon^{i}\frac{k_i}{\vert \vec{k} \vert} = 0$, on using Eq.~(\ref{pol.cond}). Thus the relevant part of the integral in Eq.~(\ref{ai.3}) is

\begin{equation}
\tilde{a}_i^{(3)}(\omega, \vec{x}) = -\frac{Mq}{16\pi^2}\frac{e^{i\omega R}}{R}\int dt \frac{r_i}{\vert \vec{r}\vert \left(|\vec{r}|+\hat{n}.\vec{r}\right)}\{ e^{i \omega \left(t - \hat{n}\vec{r}\right)} -   e^{i \omega \left(t + \vert\vec{r}\vert\right)}\} + \text{terms involving}\, n_i
\label{ai3.fin}
\end{equation}

Substituting Eq.~(\ref{as.rv}) in Eq.~(\ref{ai3.fin}), the resulting integral is of the type $I_3$ in Eq.~(\ref{int.exp}) with a finite result. Thus $\tilde{a}_i^{(3)}(\omega, \vec{x})$  does not contribute to the photon soft factor expression.

For the remaining $\tilde{a}_i(\omega, \vec{x})$ integral terms we will need to consider the action of double derivatives on $\tilde{G}_M$. Using Eq.~(\ref{as.der}) we find
\begin{align}
&\nabla_i\nabla_j\tilde{G}_M= \frac{iM}{16\pi\omega R}\left[i \frac{\omega}{|\vec{r}|+\hat{n}.\vec{r}} \left(\frac{r_j\, \hat{n}_i}{\vert \vec{r}\vert} + \hat{n}_i \hat{n}_j\right)\left( e^{i \omega \left(R - \hat{n}\vec{r}\right)} -   e^{i \omega \left(R + \vert\vec{r}\vert\right)}\right)\right. \notag\\
& \qquad \qquad - \frac{1}{\left(|\vec{r}|+\hat{n}.\vec{r}\right)^2}\left(\frac{r_i}{\vert \vec{r}\vert} + \hat{n}_i\right)\left(\frac{r_j}{\vert \vec{r}\vert} + \hat{n}_j\right)\left( e^{i \omega \left(R - \hat{n}\vec{r}\right)} -   e^{i \omega \left(R + \vert\vec{r}\vert\right)}\right) \notag\\
& \qquad \qquad \qquad + \frac{1}{|\vec{r}|+\hat{n}.\vec{r}}\left(\frac{\delta_{ij}}{\vert \vec{r}\vert} - \frac{r_ir_j}{\vert \vec{r}\vert^3}\right)\left( e^{i \omega \left(R - \hat{n}\vec{r}\right)} -   e^{i \omega \left(R + \vert\vec{r}\vert\right)}\right) \notag\\
& \qquad\qquad\qquad\qquad \left.- i\frac{\omega}{|\vec{r}|+\hat{n}.\vec{r}}\left(\frac{r_j}{\vert \vec{r}\vert} + \hat{n}_j\right)\left( \hat{n}_i e^{i \omega \left(R - \hat{n}\vec{r}\right)} + \frac{r_i}{\vert \vec{r}\vert} e^{i \omega \left(R + \vert\vec{r}\vert\right)}\right)\right]
\label{as.dder}
\end{align}
and
\begin{align}
\nabla_k\nabla_k\tilde{G}_M= \frac{M}{8\pi R |\vec{r}|} e^{i \omega \left(R +\vert\vec{r}\vert\right)}
\label{as.dder2}
\end{align}
We will now demonstrate that there is no contribution from $\tilde{a}_i^{(2)}(\omega, \vec{x})$ and $\tilde{a}_i^{(6)}(\omega, \vec{x})$. In the case of both $\tilde{a}_i^{(2)}(\omega, \vec{x})$ and $\tilde{a}_i^{(6)}(\omega, \vec{x})$, the integrand involves $v^j$ contracted with the expression in Eq.~(\ref{as.dder}). Any term involving $n_i$ does not contribute to the soft factor since $\epsilon^{i}n_i =0$. Thus using Eq.~(\ref{as.dder}) in Eq.~(\ref{ai.2}) and ignoring the $n_i$ terms, we can express $\tilde{a}_i^{(2)}(\omega, \vec{x})$ as
\begin{equation}
\tilde{a}_i^{(2)}(\omega, \vec{x}) = \tilde{a}_{i\,,I}^{(2)}(\omega, \vec{x}) + \tilde{a}_{i\,,II}^{(2)}(\omega, \vec{x}) + \tilde{a}_{i\,,III}^{(2)}(\omega, \vec{x}) + \text{terms involving}\, n_i\,,
\end{equation}
where
\begin{align}
\tilde{a}_{i\,,I}^{(2)}(\omega, \vec{x})&= - i M \frac{q}{\omega} \frac{e^{i \omega R}}{32 \pi^2 R} \int dt \left[\frac{\vec{v}.\vec{r}\, r_i}{\vert \vec{r}\vert^2 \left(\vert r\vert + \hat{n}.\vec{r}\right)^2} + \frac{\hat{n}.\vec{v}\, r_i}{\vert \vec{r}\vert \left(\vert r\vert + \hat{n}.\vec{r}\right)^2}\right]\left( e^{i \omega \left(t - \hat{n}\vec{r}\right)} -   e^{i \omega \left(t + \vert\vec{r}\vert\right)}\right)\,, \label{ai2.I}\\
\tilde{a}_{i\,,II}^{(2)}(\omega, \vec{x})&= - i M \frac{q}{\omega} \frac{e^{i \omega R}}{32 \pi^2 R} \int dt \left[\frac{\vec{v}.\vec{r}\, r_i}{\vert \vec{r}\vert^3 \left(\vert r\vert + \hat{n}.\vec{r}\right)} - \frac{v_i}{\vert \vec{r}\vert \left(\vert r\vert + \hat{n}.\vec{r}\right)}\right]\left( e^{i \omega \left(t - \hat{n}\vec{r}\right)} -   e^{i \omega \left(t + \vert\vec{r}\vert\right)}\right)\,, \label{ai2.II}\\
\tilde{a}_{i\,,III}^{(2)}(\omega, \vec{x})&=  M q \frac{e^{i \omega R}}{32 \pi^2 R} \int dt \left[\frac{\vec{v}.\vec{r}\, r_i}{\vert \vec{r}\vert^2 \left(\vert r\vert + \hat{n}.\vec{r}\right)} + \frac{\hat{n}.\vec{v}\, r_i}{\vert \vec{r}\vert \left(\vert r\vert + \hat{n}.\vec{r}\right)}\right] e^{i \omega \left(t + \vert\vec{r}\vert\right)}\,, \label{ai2.III}
\end{align}

Upon substituting Eq.~{\ref{as.rv}} in the above expressions, we identify that Eq.~(\ref{ai2.I}) and Eq.~(\ref{ai2.II}) are integrals like $I_4$ and Eq.~(\ref{ai2.III}) is like $I_5$ in Eq.~(\ref{int.exp}), using which we have
\begin{align}
\tilde{a}_{i\,,I}^{(2)}(\omega, \vec{x})&=  -M q \frac{e^{i \omega R}}{32 \pi^2 R}  \left[\frac{\beta_{+i}}{\vert \vec{\beta}_+\vert} - \frac{\beta_{-i}}{\vert \vec{\beta}_-\vert}\right] \ln \omega^{-1}  + \text{finite} = - \tilde{a}_{i\,,III}^{(2)}(\omega, \vec{x}) \,, \notag\\
\quad  & \qquad \tilde{a}_{i\,,II}^{(2)}(\omega, \vec{x}) = \text{finite} \,.
\end{align} 
Hence the evaluated expression for $\tilde{a}_{i}^{(2)}(\omega, \vec{x})$ does not contribute to the soft factor. To evaluate $\tilde{a}_{i}^{(6)}(\omega, \vec{x})$ we need the asymptotic expression for $\frac{dt}{d\sigma}$, which for large $\vert \vec{r} (t)\vert$ is
\begin{align}
\frac{dt}{d\sigma} &= \left( -g_{\mu\nu} v^{\mu}v^{\nu}\right)^{-\frac{1}{2}} = \left[ \left( 1 - \frac{M}{4 \pi \vert \vec{r}(t)\vert} \right) - \left( 1 - \frac{M}{4 \pi \vert \vec{r}(t)\vert} \right)^{-1} \vec{v}(t)^2 \right]^{-\frac{1}{2}} \notag\\
& = \frac{1}{\sqrt{1 - \vec{v}(t)^2}} \left( 1 + \frac{M}{8 \pi \vert \vec{r}(t)\vert}\frac{1 + \vec{v}(t)^2}{1 - \vec{v}(t)^2}\right) + \mathcal{O}\left(\phi^2\right)\notag\\ 
&= \frac{1}{\sqrt{1 - \vec{\beta}_{\pm}^2}} \left( 1 \pm \frac{M}{8 \pi \vert \vec{\beta}_{\pm}\vert}\frac{1 + \vec{\beta}_{\pm}^2}{1 - \vec{\beta}_{\pm}^2}\right) + \mathcal{O}\left(t^{-2}\right)
\label{as.dtds}
\end{align}
As in the case $\tilde{a}_{i}^{(2)}(\omega, \vec{x})$, we express $\tilde{a}_{i}^{(6)}(\omega, \vec{x})$ as
\begin{equation}
\tilde{a}_i^{(6)}(\omega, \vec{x}) = \tilde{a}_{i\,,I}^{(6)}(\omega, \vec{x}) + \tilde{a}_{i\,,II}^{(6)}(\omega, \vec{x}) + \tilde{a}_{i\,,III}^{(6)}(\omega, \vec{x}) + \text{terms involving}\, n_i\,,
\end{equation}
where
\begin{align}
\tilde{a}_{i\,,I}^{(6)}(\omega, \vec{x})&= i Q \frac{m}{\omega} \frac{e^{i \omega R}}{4 \pi R} \int dt\frac{dt}{d\sigma} \left[\frac{\vec{v}.\vec{r}\, r_i}{\vert \vec{r}\vert^2 \left(\vert r\vert + \hat{n}.\vec{r}\right)^2} + \frac{\hat{n}.\vec{v}\, r_i}{\vert \vec{r}\vert \left(\vert r\vert + \hat{n}.\vec{r}\right)^2}\right]\left( e^{i \omega \left(t - \hat{n}\vec{r}\right)} -   e^{i \omega \left(t + \vert\vec{r}\vert\right)}\right)\,, \label{ai6.I}\\
\tilde{a}_{i\,,II}^{(6)}(\omega, \vec{x})&= i Q \frac{m}{\omega} \frac{e^{i \omega R}}{4 \pi R} \int dt \frac{dt}{d\sigma} \left[\frac{\vec{v}.\vec{r}\, r_i}{\vert \vec{r}\vert^3 \left(\vert r\vert + \hat{n}.\vec{r}\right)} - \frac{v_i}{\vert \vec{r}\vert \left(\vert r\vert + \hat{n}.\vec{r}\right)}\right]\left( e^{i \omega \left(t - \hat{n}\vec{r}\right)} -   e^{i \omega \left(t + \vert\vec{r}\vert\right)}\right)\,, \label{ai6.II}\\
\tilde{a}_{i\,,III}^{(6)}(\omega, \vec{x})&= - Q m \frac{e^{i \omega R}}{4 \pi R} \int dt \frac{dt}{d\sigma} \left[\frac{\vec{v}.\vec{r}\, r_i}{\vert \vec{r}\vert^2 \left(\vert r\vert + \hat{n}.\vec{r}\right)} + \frac{\hat{n}.\vec{v}\, r_i}{\vert \vec{r}\vert \left(\vert r\vert + \hat{n}.\vec{r}\right)}\right] e^{i \omega \left(t + \vert\vec{r}\vert\right)}\,, \label{ai6.III}
\end{align}
Using the expansions in Eq.~(\ref{as.rv}) and Eq.~(\ref{as.dtds}), with the expressions for $I_4$ and $I_5$ in Eq.~(\ref{int.exp}), we find
\begin{align}
\tilde{a}_{i\,,I}^{(6)}(\omega, \vec{x})&=  Q m \frac{e^{i \omega R}}{4 \pi R}  \left[\frac{\beta_{+i}}{\vert \vec{\beta}_+\vert \sqrt{1 - \vec{\beta}_+^2}} - \frac{\beta_{-i}}{\vert \vec{\beta}_-\vert\sqrt{1 - \vec{\beta}_-^2}}\right] \ln \omega^{-1}  + \text{finite} = - \tilde{a}_{i\,,III}^{(2)}(\omega, \vec{x}) \,, \notag\\
\quad  & \qquad \tilde{a}_{i\,,II}^{(2)}(\omega, \vec{x}) = \text{finite} \,.
\end{align} 
Thus $\tilde{a}_{i}^{(6)}(\omega, \vec{x})$ also does not contribute to the electromagnetic soft factor. 

The final integral to consider in the electromagnetic solution is that for $\tilde{a}_{i}^{(5)}(\omega, \vec{x})$. Substituting Eq.~(\ref{as.dder2}) in Eq.~(\ref{ai.5}) we get
\begin{align}
\tilde{a}_{i\,}^{(5)}(\omega, \vec{x})&= -\frac{M q}{32 \pi^2}\frac{e^{i \omega R}}{R} \int dt \frac{v_i}{\vert \vec{r} \vert} e^{i \omega \left(t + \vert\vec{r}\vert\right)}
\label{ai5.sub}
\end{align}

On substituting Eq.~(\ref{as.rv}) in the integrand of $\tilde{a}_{i}^{(5)}(\omega, \vec{x})$ and using the expression for $I_5$ in Eq.~(\ref{int.exp}), we find
\begin{equation}
\tilde{a}_{i}^{(5)}(\omega, \vec{x}) = - M q \frac{e^{i \omega R}}{32 \pi^2 R} \left(\frac{\beta_{+i}}{\vert \vec{\beta}_+ \vert} - \frac{\beta_{-i}}{\vert \vec{\beta}_- \vert}\right) \ln \omega^{-1} + \text{finite} \,.
\label{ai52.fin}
\end{equation}
By using Eq.~(\ref{ai52.fin}) in Eq.~(\ref{sfact4.em}), we find the following contribution of this term to the soft factor 
\begin{equation}
i \frac{4 \pi R}{e^{i \omega R}} \epsilon^{i}\tilde{a}_{i}^{(5)}(\omega, \vec{x}) = - i \frac{M}{8 \pi} q \ln \omega^{-1} \left(\frac{\vec{\epsilon}. \vec{\beta}_+}{\vert \vec{\beta}_+\vert} + \frac{\vec{\epsilon}. \vec{\beta}_-}{\vert \vec{\beta}_-\vert}\right)
\label{ai52.soft}
\end{equation}
This is a subleading contribution, which cancels out the $M$ terms in the subleading contribution of Eq.~(\ref{ai1.soft}). From the combination of all non-vanishing contributions to the soft factor, i.e.  Eq.~(\ref{ai1.soft}), Eq.~(\ref{ai4.soft}) and Eq.~(\ref{ai52.soft}), we find
\begin{align}
&S_{\text{em}} =i \frac{4 \pi R}{e^{i \omega R}} \epsilon^{i}\left(\tilde{a}_i^{(1)}(\omega, \vec{x}) + \tilde{a}_{i}^{(4)}(\omega, \vec{x}) + \tilde{a}_i^{(5)}(\omega, \vec{x})\right) \notag\\
& \qquad = - \frac{q}{\omega} \left(\frac{\vec{\epsilon}. \vec{\beta}_+}{1 - \hat{n}.\vec{\beta}_+} - \frac{\vec{\epsilon}. \vec{\beta}_-}{1 - \hat{n}.\vec{\beta}_-}\right) - i q \ln \omega^{-1} \left(\frac{C_+}{1 - \hat{n}.\vec{\beta}_+}\vec{\epsilon}. \vec{\beta}_+ - \frac{C_-}{1 - \hat{n}.\vec{\beta}_-}\vec{\epsilon}. \vec{\beta}_-\right) \notag\\
&\qquad  \qquad - i q \frac{M}{4 \pi} \ln\left(\omega R\right) \left(\frac{\vec{\epsilon}. \vec{\beta}_+}{1-\hat{n}.\vec{\beta}_+} - \frac{\vec{\epsilon}. \vec{\beta}_-}{1-\hat{n}.\vec{\beta}_-}\right)
\label{ai.soft1}
\end{align}
The second line of Eq.~(\ref{ai.soft1}) agrees with the expression for the soft factor in Eq.~(\ref{sem.pp}), while the third line in Eq.~(\ref{ai.soft1}) agrees with the predicted phase contribution to the soft factor in Eq.~(\ref{semc.pp}) resulting from the backscattering of soft photons due to the gravitational potential of the RN black hole.

\subsection{Gravitational soft factors and phases on the RN spacetime}\label{gravisoft}

We will now carry out the analysis of the previous subsection on the solutions for gravitational perturbations in Eq.~(\ref{hij.fin}). Using Eq.~(\ref{soft.gr}), Eq.~(\ref{soft.norm}) and Eq.~(\ref{pol.cond}), we find the following expression for $S_{\text{gr}}$ in $D=4$ dimensions
\begin{equation}
S_{\text{gr}} = i \frac{4 \pi R}{e^{i \omega R}} \epsilon^{ij}\tilde{e}_{ij} (\omega, \vec{x}) \,,\label{sfact4.gr}
\end{equation}
As in the electromagnetic case, we let $r^0 = t$ and $x >> r(t)$ by substituting Eq.~(\ref{gfn.lim}) in Eq.~(\ref{hij.fin}). The resulting expression for Eq.~(\ref{hij.fin}) can be expressed as
\begin{equation}
\tilde{e}_{ij}(\omega, \vec{x}) = \tilde{e}^{(1)}_{ij}(\omega, \vec{x}) + \tilde{e}^{(2)}_{ij}(\omega, \vec{x}) + \tilde{e}^{(3)}_{ij}(\omega, \vec{x})+ \tilde{e}^{(4)}_{ij}(\omega, \vec{x}) + \tilde{e}^{(5)}_{ij}(\omega, \vec{x}) + \tilde e^{(6)}_{ij}(\omega, \vec{x})+ \tilde e^{(7)}_{ij}(\omega, \vec{x})\,,
\end{equation}
where
\begin{align} 
\tilde{e}^{(1)}_{ij}(\omega, \vec{x}) 
&=\frac{m~e^{i\omega R}}{ 4\, \pi\, R} \int \frac{dt}{\left(1+2\Phi(\vec{r}(t))\right)}\frac{dt}{d\sigma}\, v_i v_j \, 
e^{i\omega (t-\hat{n}.\vec{r}(t))} +\text{boundary terms}
 \, ,\label{eij.1}\\
\tilde e^{(2)}_{ij}(\omega, \vec{x}) &= \frac{m}{2\pi} \int dt \frac{dt}{d\sigma}\, e^{i\omega t}(1+\vec{v}^2) \, \left(\nabla_i\nabla_j -\frac{1}{2} \delta_{ij} \, \nabla_k\nabla_k\right)\tilde{G}_M\left(\omega,\vec{x},\vec{r}\right), \label{eij.2}\\
\tilde e^{(3)}_{ij}(\omega, \vec{x})&= -i\, \frac{M m}{16\, \pi^2} \, \omega \, \frac{e^{i\omega R}}{R}\int dt \frac{dt}{d\sigma}\, v_i\, v_j \,\bigg\{ \ln \frac{|\vec{r}\,'|+\hat{n}.\vec{r}\,'}{ R} \, e^{i\omega (t - \hat{n}.\vec{r}\,')}+ \int_{|\vec{r}\,'|+\hat{n}.\vec{r}\,'}^\infty \frac{du}{u} e^{i\omega (t - \hat{n}.\vec{r}\,'+u)}\bigg\}, \label{eij.3}\\
\tilde e^{(4)}_{ij}(\omega, \vec{x}) &= \frac{i\omega m}{\pi} \int dt \frac{dt}{d\sigma}\, e^{i\omega t}\left(v_i\nabla_j + v_j \nabla_i \right)\,\tilde{G}_M\left(\omega,\vec{x},\vec{r}\right), \label{eij.4}\\ 
\tilde e^{(5)}_{ij}(\omega, \vec{x}) &= -i\omega\frac{qQ}{2\pi M}\int dt e^{i\omega t}\delta_{ij}v^k\nabla_k\tilde{G}_M\left(\omega,\vec{x},\vec{r}\right), \label{eij.5}\\
\tilde e^{(6)}_{ij}(\omega, \vec{x}) &=i\frac{qQ}{\pi M} \int dte^{i\omega t}\left(\nabla_i\nabla_j -\frac{1}{2} \delta_{ij} \, \nabla_k\nabla_k\right)\tilde{G}_M\left(\omega,\vec{x},\vec{r}\right),
\label{eij.6}\\
\tilde e^{(7)}_{ij}(\omega, \vec{x}) &= i\omega\frac{qQ}{2\pi M} \int dt e^{i\omega t}\left(v_i\nabla_j + v_j \nabla_i \right)\tilde{G}_M\left(\omega,\vec{x},\vec{r}\right)\label{eij.7}
\end{align}
The $\tilde e^{(1)}_{ij}(\omega, \vec{x})\, \cdots \tilde e^{(4)}_{ij}(\omega, \vec{x})$ contributions are those of the Schwarzschild spacetime. In the $\tilde e^{(1)}_{ij}(\omega, \vec{x})$ integral in Eq.~(\ref{eij.1}), we substitute Eq.~(\ref{as.rv}), Eq.~(\ref{as.pot}) and Eq.~(\ref{as.dtds}). After carrying out an integration by parts, the resulting integral is of the $I_1$ type in Eq.~(\ref{int.exp}) with the result
\begin{align}
\tilde e^{(1)}_{ij}(\omega, \vec{x}) &= i\,{m\over 4\pi\,\omega } {e^{i\omega R}\over R} 
\left\{ {1\over 1 - \hat n.\vec \beta_+} \, {1\over \sqrt{1-\vec \beta_+^2}}\, \beta_{+ i} 
\beta_{+ j} - {1\over 1 - \hat n.\vec \beta_-} \, {1\over \sqrt{1-\vec \beta_-^2}}\, \beta_{- i} 
\beta_{- j}\right\}\nonumber \\
& - {m\over 4\pi\, R} e^{i\omega R}  \ln \omega^{-1} \left[ 
{1\over \sqrt{1-\vec \beta_+^2}}\, \beta_{+ i} 
\beta_{+ j} \left\{C_+ {1\over 1-\hat n.\vec\beta_+} - 
{ M_0\over 8\, \pi\, |\vec\beta_+|} \, {3-\vec\beta_+^2\over 1-\vec\beta_+^2}
+C_+ {1\over 1 -\vec\beta_+^2} 
\right\} \right.\nonumber \\ &
\left. - {1\over \sqrt{1-\vec \beta_-^2}}\, \beta_{- i} 
\beta_{- j} \left\{C_- {1\over 1-\hat n.\vec\beta_-} + 
{M_0\over 8\pi |\vec\beta_-|} \, {3-\vec\beta_-^2\over 1-\vec\beta_-^2}
+C_- {1\over 1 -\vec\beta_-^2} 
\right\}
\right]\, ,
\label{eij1.fin}
\end{align}
We next consider the integral for $\tilde e^{(3)}_{ij}(\omega, \vec{x})$ in Eq.~(\ref{eij.3}) and substitute Eq.~(\ref{as.rv}) and Eq.~(\ref{as.dtds}). This integral also requires an integration by parts, leading to an integral of the $I_2$ form in Eq.~(\ref{int.exp}) with the result
\begin{align}
\tilde e^{(3)}_{ij}(\omega, \vec{x})=
- {M m\over 16\, \pi^2}  \, \ln(\omega R) \, {e^{i\omega R}\over R} \, 
\left\{ {1\over 1 - \hat n.\vec \beta_+} \, {1\over \sqrt{1-\vec \beta_+^2}}\, \beta_{+ i} 
\beta_{+ j} - {1\over 1 - \hat n.\vec \beta_-} \, {1\over \sqrt{1-\vec \beta_-^2}}\, \beta_{- i} 
\beta_{- j}\right\}\, .
\label{eij3.fin}
\end{align}

The integrals for $\tilde e^{(2)}_{ij}(\omega, \vec{x})$ and $\tilde e^{(6)}_{ij}(\omega, \vec{x})$ in Eq.~(\ref{eij.2}) and Eq.~(\ref{eij.6}) respectively are evaluated in an identical manner. In these integrals, terms which involve $\delta_{ij}$ do not contribute to the soft factor as the polarization tensor satisfies $\epsilon^{ij}\delta_{ij} = \epsilon^i_i = 0$ using Eq.~(\ref{pol.cond}). Furthermore, from Eq.~(\ref{as.dder}), we also note that for the $\partial_i \partial_j$ contribution, only those terms which do not involve $\hat{n}$ are relevant to the soft factor. This follows from the condition $\epsilon^{ij}n_i = \epsilon^{ij}\frac{k_i}{\vert \vec{k}\vert} =  0$. 

Thus in the case of $\tilde e^{(6)}_{ij}(\omega, \vec{x})$ in Eq.~(\ref{eij.6}, we find the following integral on using Eq.~(\ref{as.dder})
\begin{align}  
\tilde e^{(6)}_{ij}(\omega, \vec{x}) &= Q q \frac{e^{i \omega R}}{16 \pi^2 R} \int dt \frac{r_i r_j}{\vert \vec{r} \vert^2 \left(\vert \vec{r} \vert + \hat{n}.\vec{r}\right)} e^{i \omega (t+ \vert \vec{r} \vert)} \notag\\ 
& \hspace{-3em}- i Q \frac{q}{\omega} \frac{e^{i \omega R}}{16 \pi^2 R} \int dt \left[\left(\frac{r_i r_j}{\left(\vert \vec{r} \vert \left(\vert \vec{r} \vert + \hat{n}.\vec{r}\right)\right)^2} + \frac{r_i r_j}{\vert \vec{r} \vert^3 \left(\vert \vec{r} \vert + \hat{n}.\vec{r}\right)}\right) \left(e^{i \omega(t - \hat{n}.\vec{r})} - e^{i \omega(t + \vert \vec{r} \vert )}\right)\right] \notag\\
& \qquad\qquad\qquad\qquad\qquad\qquad + \text{terms involving} \, n_i \, \text{or}\, n_j 
\label{eij6.int}
\end{align}
On susbstituting Eq.~(\ref{as.rv}) in Eq.~(\ref{eij6.fin}), we find that the integral in the first line is of the $I_5$ type and the integral in the second line is of the $I_4$ type of Eq.~(\ref{int.exp}). Using the corresponding results for these integrals, we find that Eq.~(\ref{eij6.fin}) evaluates to
\begin{align}
\tilde e^{(6)}_{ij}(\omega, \vec{x}) &= Q q \frac{e^{i \omega R}}{16 \pi^2 R}\ln \omega^{-1}\left(\frac{\beta_{+i}\beta_{+j}}
{\vert \vec\beta_+\vert^2 \left(\vert \vec{\beta}_+\vert + \hat{n}\vec\beta_+\right)} + \frac{\beta_{-i}\beta_{-j}}
{\vert \vec\beta_-\vert^2 \left(\vert \vec{\beta}_-\vert - \hat{n}\vec\beta_-\right)}\right) \notag\\ 
& \hspace{-3em} -Q q \frac{e^{i \omega R}}{16 \pi^2 R}\ln \omega^{-1}\left[\left(\frac{\beta_{+i}\beta_{+j}}
{\vert \vec\beta_+\vert^2 \left(\vert \vec{\beta}_+\vert + \hat{n}\vec\beta_+\right)} + \frac{\beta_{-i}\beta_{-j}}
{\vert \vec\beta_-\vert^2 \left(\vert \vec{\beta}_-\vert - \hat{n}\vec\beta_-\right)}\right) + \frac{\beta_{+i}\beta_{+j}}{\vert \vec\beta_+\vert^3} + \frac{\beta_{-i}\beta_{-j}}{\vert \vec\beta_-\vert^3}\right] + \text{finite} 
\end{align}
Hence
\begin{equation}
\tilde e^{(6)}_{ij}(\omega, \vec{x}) =  -Q q \frac{e^{i \omega R}}{16 \pi^2 R}\ln \omega^{-1}\left[\frac{\beta_{+i}\beta_{+j}}{\vert \vec\beta_+\vert^3} + \frac{\beta_{-i}\beta_{-j}}{\vert \vec\beta_-\vert^3}\right] + \text{finite}
\label{eij6.fin}
\end{equation}

Likewise, in the case of $\tilde e^{(2)}_{ij}(\omega, \vec{x})$ in Eq.~(\ref{eij.2}, we find on using Eq.~(\ref{as.dder})
\begin{align}  
\tilde e^{(2)}_{ij}(\omega, \vec{x}) &= M m \frac{e^{i \omega R}}{32 \pi^2 R} \int dt \frac{dt}{d\sigma} \frac{r_i r_j}{\vert \vec{r} \vert^2 \left(\vert \vec{r} \vert + \hat{n}.\vec{r}\right)} e^{i \omega (t+ \vert \vec{r} \vert)} \notag\\ 
& \hspace{-3em}- i M \frac{m}{\omega} \frac{e^{i \omega R}}{32 \pi^2 R} \int dt \frac{dt}{d\sigma} \left[\left(\frac{r_i r_j}{\left(\vert \vec{r} \vert \left(\vert \vec{r} \vert + \hat{n}.\vec{r}\right)\right)^2} + \frac{r_i r_j}{\vert \vec{r} \vert^3 \left(\vert \vec{r} \vert + \hat{n}.\vec{r}\right)}\right) \left(e^{i \omega(t - \hat{n}.\vec{r})} - e^{i \omega(t + \vert \vec{r} \vert )}\right)\right] \notag\\
& \qquad\qquad\qquad\qquad\qquad\qquad + \text{terms involving} \, n_i \, \text{or}\, n_j \,.
\label{eij2.int}
\end{align}
The integral in first line of Eq.~(\ref{eij2.int}) is of the $I_5$ type and the integral in the second line of Eq.~(\ref{eij2.int}) is of the $I_4$ type given in Eq.~(\ref{int.exp}), hence providing the result

\begin{equation}
\tilde e^{(2)}_{ij}(\omega, \vec x) =- M m \frac{e^{i \omega R}}{32 \pi^2 R} \ln\omega^{-1} \left[{(1+\vec\beta_+^2) \beta_{+i}\beta_{+j}\over
|\vec\beta_+|^3\, \sqrt{1-\vec\beta_+^2}} + 
{(1+\vec\beta_ -^2) \beta_{ -i}\beta_{ -j}\over
|\vec\beta_ -|^3\, \sqrt{1-\vec\beta_ -^2}}\right]\,. 
\label{eij2.fin}
\end{equation}

All remaining integrals do not contribute to the soft factor. The $\tilde e^{(5)}_{ij}(\omega, \vec{x})$ integral in Eq.~(\ref{eij.5} does not contribute since $\epsilon^{ij}\delta_{ij} = \epsilon^i_i =0$. The integrals for $\tilde e^{(4)}_{ij}(\omega, \vec{x})$ in Eq.~(\ref{eij.4}) and $\tilde e^{(7)}_{ij}(\omega, \vec{x})$ in Eq.~(\ref{eij.7}) are of a similar type. In both integrals, we substitute Eq.~(\ref{as.der}) and seek the possible contributions to the soft factor. All terms involving $\hat{n}$ do not contribute, since $\epsilon^{ij}n_i = \epsilon^{ij}\frac{k_i}{\vert \vec{k}\vert} =  0$ from Eq.~(\ref{pol.cond}). Thus the relevant terms for the soft factor in Eq.~(\ref{eij.4}) and Eq.~(\ref{eij.7}) on using Eq.~(\ref{as.der}) are
\begin{align}
\tilde e^{(4)}_{ij}(\omega, \vec{x}) & = - M m \frac{e^{i\omega R}}{16 \pi^2 R} \int dt \frac{dt}{d\sigma}\frac{\left(v_i r_j + v_j r_i\right)}{\vert\vec r \vert \left(\vert\vec r \vert + \hat{n}.\vec{r}\right)} \left(e^{i \omega(t - \hat{n}.\vec{r})} - e^{i \omega(t + \vert \vec{r} \vert )}\right) + \text{terms involving}\, n_j\, \text{or}\,n_i \,,\label{eij4.fin}\\
\tilde e^{(7)}_{ij}(\omega, \vec{x}) & = - Q q \frac{e^{i\omega R}}{32 \pi^2 R} \int dt \frac{ \left(v_i r_j + v_j r_i\right)}{\vert\vec r \vert \left(\vert\vec r \vert + \hat{n}.\vec{r}\right)} \left(e^{i \omega(t - \hat{n}.\vec{r})} - e^{i \omega(t + \vert \vec{r} \vert )}\right) + \text{terms involving}\, n_j\, \text{or}\, n_i \,,
\label{eij7.fin}
\end{align} 
On substituting Eq.~(\ref{as.rv}) and Eq.~(\ref{as.dtds}) in Eq.~(\ref{eij4.fin}), and Eq.~(\ref{as.rv}) in Eq.~(\ref{eij7.fin}), we find that the above integrals are both of the $I_3$ type in Eq.~(\ref{int.exp}) and hence do not contribute any $\omega^{-1}$ or $\ln \omega^{-1}$ terms. 

Thus the only terms which do contribute to the gravitation soft factor are from Eq.~(\ref{eij1.fin}), Eq.~(\ref{eij3.fin}), Eq.~(\ref{eij6.fin}) and Eq.~(\ref{eij2.fin}), which on substituting in Eq.~(\ref{sfact4.gr}) gives

\begin{align}
S_{\text{gr}} &= i \frac{4 \pi R}{e^{i \omega R}} \epsilon^{ij}\tilde{e}_{ij} (\omega, \vec{x})\notag\\
&= - \frac{m}{\omega} \epsilon^{ij}\left[\frac{1}{1 - \hat{n}.\vec{\beta}_+}\frac{1}{\sqrt{1 - \vec{\beta}^2_+}}\beta_{+i} \beta_{+j} - \frac{1}{1 - \hat{n}.\vec{\beta}_-}\frac{1}{\sqrt{1 - \vec{\beta}^2_-}}\beta_{-i} \beta_{-j}\right] \notag\\
&\quad -im\ln \omega^{-1}\epsilon^{ij}\,\left[{1\over \sqrt{1-\vec \beta_+^2}}\, \left\{C_+ \left({1\over 1-\hat n.\vec\beta_+} +{1\over 1 -\vec\beta_+^2}\right)-{ M\over 8\, \pi\, |\vec\beta_+|^3}\, {3\vec\beta_+^2 -1 \over 1-\vec \beta_+^2}\right\}\beta_{+ i} 
\beta_{+ j}  \right.\nonumber \\ &
\left. \qquad \qquad \qquad \qquad \qquad  - {1\over \sqrt{1-\vec \beta_-^2}}\,  \left\{C_- \left({1\over 1-\hat n.\vec\beta_-}+{1\over 1 -\vec\beta_-^2} \right) + 
{M\over 8\, \pi\, |\vec\beta_-|^3} {3\vec\beta_-^2-1\over 1-\vec\beta_-^2}
\right\}\beta_{- i} \beta_{- j} \right] \notag\\
& \qquad - i m \frac{M}{4 \pi} \ln \left(R \omega\right) \epsilon^{ij}\left[\frac{1}{1 - \hat{n}.\vec{\beta}_+}\frac{1}{\sqrt{1 - \vec{\beta}^2_+}}\beta_{+i} \beta_{+j} - \frac{1}{1 - \hat{n}.\vec{\beta}_-}\frac{1}{\sqrt{1 - \vec{\beta}^2_-}}\beta_{-i} \beta_{-j}\right] \notag\\
& \quad \qquad -i \frac{qQ}{4\pi}\ln \omega^{-1}\epsilon^{ij}\,\left[ \frac{\beta_{+ i}\beta_{+ j}}{ |\vec\beta_+|^3}+\frac{\beta_{- i}\beta_{- j}}{ |\vec\beta_-|^3}\right] + \text{finite}\label{eij.soft1}
\end{align}
The leading contribution in the above expression agrees with Eq.~(\ref{sgr.pp}) and the phase correction agrees with Eq.~(\ref{sgrc.pp}). To complete the agreement with the subleading term in Eq.~(\ref{sgr.pp}), we need the expression for $C_{\pm}$. This can be determined by considering the energy of the probe particle from the point particle action in Eq.~(\ref{act.pp})
\begin{align}
- E &= \frac{\delta S_{pp}}{\delta(\frac{dt}{d\sigma})}\notag\\
&=-m\vert g_{00}\vert
\frac{dt}{d\sigma}+\frac{q}{4\pi}A_0
\label{eom.pp}
\end{align}
Using Eq.~(\ref{rn.gfa}), Eq.~(\ref{as.rv}), Eq.~(\ref{as.dtds}) and Eq.~(\ref{as.pot}) in the last line of Eq.~(\ref{eom.pp}), we find
\begin{align}
&- m\vert g_{00}\vert \frac{dt}{d\sigma}+\frac{q}{4\pi}A_0\notag\\
&= -\frac{m}{\sqrt{1 - \vec{\beta}_{\pm}^2}}\left[1 - \frac{1}{t}\left( \frac{C_{\pm}\vec{\beta}_{\pm}^2}{1 - \vec{\beta}_{\pm}^2} \mp \frac{M}{8 \pi \vert \vec{\beta}_{\pm}\vert} \left(\frac{3\vec\beta_{\pm}^2-1}{1 - \vec{\beta}_{\pm}^2}\right)\right)\right] \pm \frac{1}{t}\frac{q Q}{\vert \vec{\beta}_{\pm}\vert} + \mathcal{O}\left(t^{-2}\right)
\label{en.pp}
\end{align}
We get the desired expression relating $C_{\pm}$ with $M$ and $Q$ from Eq.~(\ref{en.pp}) by setting the $\frac{1}{t}$ coefficient to vanish
\begin{align}
C_{\pm}=\pm{M\over 8\, \pi\, |\vec\beta_{\pm}|^3} (3\vec\beta_{\pm}^2-1)\mp\frac{qQ}{4\pi m  |\vec\beta_{\pm}|^3}(1-\vec\beta_{\pm}^2)^{3/2}\,.
\label{c.pp}
\end{align}
Substituting this expression in (\ref{eij.soft1}) gives
\begin{align} 
S_{\text{gr}} =& - \frac{m}{\omega} \epsilon^{ij}\left[\frac{1}{1 - \hat{n}.\vec{\beta}_+}\frac{1}{\sqrt{1 - \vec{\beta}^2_+}}\beta_{+i} \beta_{+j} - \frac{1}{1 - \hat{n}.\vec{\beta}_-}\frac{1}{\sqrt{1 - \vec{\beta}^2_-}}\beta_{-i} \beta_{-j}\right] \notag\\
&-i m \ln \omega^{-1} \epsilon^{ij} \left[C_+\frac{1}{1 - \hat{n}.\vec{\beta}_+}\frac{1}{\sqrt{1 - \vec{\beta}^2_+}}\beta_{+i} \beta_{+j} - C_-\frac{1}{1 - \hat{n}.\vec{\beta}_-}\frac{1}{\sqrt{1 - \vec{\beta}^2_-}}\beta_{-i} \beta_{-j}\right] \notag\\
&- i m \frac{M}{4 \pi} \ln \left(R \omega\right) \epsilon^{ij}\left[\frac{1}{1 - \hat{n}.\vec{\beta}_+}\frac{1}{\sqrt{1 - \vec{\beta}^2_+}}\beta_{+i} \beta_{+j} - \frac{1}{1 - \hat{n}.\vec{\beta}_-}\frac{1}{\sqrt{1 - \vec{\beta}^2_-}}\beta_{-i} \beta_{-j}\right]
\label{eij.soft}
\end{align}

The first two lines are now in complete agreement with the soft factor in Eq.~(\ref{sgr.pp}) and the last line of Eq.~(\ref{eij.soft}) has the phase correction given in Eq.~(\ref{sgrc.pp}).

\subsection{Memory and tail effects from soft factors}\label{gravemwav}

We have verified that the perturbation solutions in Sec.~\ref{classical} provide the predicted soft factors resulting from the classical limit of photon and graviton soft theorems. 
As mentioned, the photon and graviton soft factor contributions considered in this section are universal. In addition, the soft factor results in the previous subsections involved an independent sum over the incoming and outgoing state of the probe particle. Hence the soft factor results in Eq.~(\ref{eij.soft}) and Eq.~(\ref{ai.soft1}) can be generalized to the case where we only consider light particles in the outgoing state and no light particles in the incoming state. 

This was the approach taken in~\cite{Laddha:2018vbn}, where the graviton soft factor resulting from the probe scattering on the $D=4$ Schwarzschild spacetime~\cite{Laddha:2018myi} was generalized to the case of several light particles in the outgoing state and no incoming light particles. With this assumption on light particle states, their analysis considered the effect of the overall phase and logarithmic terms in the soft factor on the expression for the transverse and traceless components of gravitational waves. It was demonstrated that memory effect in gravitational waves now involves a tail, with contributions arising from the logarithmic terms and phase correction present in the subleading soft factor.
The resulting solution in frequency space takes the form
\begin{equation}
\tilde{e}_{ij}^{TT}\left(\omega\,, \vec{x}\right) = i \omega^{-1} A_{ij} + \ln \omega^{-1} B_{ij} + \text{finite}\,, \label{sl.exp}
\end{equation}
where $A_{ij}$ is the leading memory contribution and $B_{ij}$ is the subleading tail term with expressions 
\begin{align}
A_{ij} & = \frac{2 G}{R} \sum_{a} m_a \frac{1}{1- \hat{n}.\vec\beta_{a}}\frac{1}{\sqrt{1- \vec\beta_{a}^2}} \left(\beta_{a i}\beta_{a j}\right)^{TT} \,, \label{sl.mem} \\
B_{ij} & = \frac{2 G^2 M}{R} \sum_{a} m_a \frac{1}{1- \hat{n}.\vec\beta_{a}}\frac{1}{\sqrt{1- \vec\beta_{a}^2}}\frac{1- 3\vec\beta_{a}^2 + 2 \vert \vec\beta_{a} \vert^3}{\vert \vec\beta_{a} \vert^3} \left(\beta_{a i}\beta_{a j}\right)^{TT}\,. \label{sl.tail}
\end{align}
In Eq.~(\ref{sl.mem}) and Eq.~(\ref{sl.tail}), the sum over all outgoing soft particles is denoted by `$a$' and the superscript `$TT$' refers to the transverse and traceless components. In addition, the dependence on Newton's constant $G$ has been made explicit which reveals that the leading memory contribution $A_{ij}$ dominates the subleading tail term $B_{ij}$. 

The above result and the scattering approximation involved are realized in events where the total mass of the heavy scatterer before and after the scattering event are comparable, and far greater than the energy of the emitted light particles. Examples of such processes in particular involve binary black hole and neutron star mergers. The emitted light particles can be considered in the ultra-relativistic limit $\vert \vec\beta_a\vert \to 1$. A strict massless limit of the outgoing light states, i.e. $m_a \to 0$, cannot be taken if the results are based on the probe-scatterer approximation as considered in this paper and in~\cite{Peters:1970mx}. Nevertheless, the result in the ultra-relativistic limit can be used to infer implications for gravitational waves emitted from binary black hole events. As can be noted from Eq.~(\ref{sl.tail}), in this limit $B_{ij} \to 0$ and there exists no non-linear tail effect. 

We will now directly make use of the soft factor expressions for $S_{\text{gr}}$ in Eq.~(\ref{eij.soft}) and $S_{\text{em}}$ in Eq.~(\ref{ai.soft1}) to determine their implications on the memory and tail terms. We will reinstate the dependence on $G$ in these equations and throughout this subsection. The soft factor results in the probe-scatterer approximation can be generalized to the case where there only exist several light particles in the outgoing state. Expressing $\ln (R \omega) = \ln R - \ln \omega^{-1}$, from $S_{\text{gr}}$ in Eq.~(\ref{eij.soft}) we find

\begin{align}
S_{\text{gr}} &\xrightarrow{\text{single outgoing probe}} - 8 \pi G \frac{m}{\omega} \epsilon^{ij}\left(\frac{1}{1 - \hat{n}.\vec{\beta}_+}\frac{1}{\sqrt{1 - \vec{\beta}^2_+}}\beta_{+i} \beta_{+j}\right)\notag\\
& \qquad \qquad  -i 8 \pi G m \ln \omega^{-1} \epsilon^{ij} \left[\left(C_+ - 2 G M\right)\frac{1}{1 - \hat{n}.\vec{\beta}_+}\frac{1}{\sqrt{1 - \vec{\beta}^2_+}}\beta_{+i} \beta_{+j}\right] + \text{finite}\notag\\
&\xrightarrow{\text{light outgoing particles}}  - 8 \pi G \left(\frac{\epsilon^{ij}}{\omega} \sum_{a} m_a \frac{1}{1 - \hat{n}.\vec{\beta}_a}\frac{1}{\sqrt{1 - \vec{\beta}^2_a}}\beta_{ai} \beta_{aj}\right) \notag\\
& \qquad \qquad  - 8 \pi G \left( i \ln \omega^{-1} \epsilon^{ij} \sum_{a} m_a \left(C_a - 2 G M\right)\frac{1}{1 - \hat{n}.\vec{\beta}_a}\frac{1}{\sqrt{1 - \vec{\beta}^2_a}}\beta_{a i} \beta_{a j}\right) + \text{finite}\,.\notag\\
\label{eij.msoft1} 
\end{align}
In the first generalization of Eq.~(\ref{eij.msoft1}), we ignored the incoming state contribution present in Eq.~(\ref{eij.soft}), while in the second generalization we changed the outgoing state into a sum over several light particles. Denoting the result for the graviton soft factor in this case by $\tilde{S}_{\text{gr}}$ and reinstating the dependence on $G$, we have
\begin{align}
\tilde{S}_{\text{gr}} &= - 8 \pi G \left(\frac{\epsilon^{ij}}{\omega} \sum_{a} m_a \frac{1}{1 - \hat{n}.\vec{\beta}_a}\frac{1}{\sqrt{1 - \vec{\beta}^2_a}}\beta_{ai} \beta_{aj}\right)\notag\\
& \qquad \qquad  - 8 \pi G \left( i \ln \omega^{-1} \epsilon^{ij} \sum_{a} m_a \left(C_a - 2 G M\right)\frac{1}{1 - \hat{n}.\vec{\beta}_a}\frac{1}{\sqrt{1 - \vec{\beta}^2_a}}\beta_{a i} \beta_{a j}\right) + \text{finite}\,.
\label{eij.msoft}
\end{align}
Likewise, by repeating the above steps for the photon soft factor $S_{\text{em}}$ in Eq.~(\ref{ai.soft1}), we find the following expression for $\tilde{S}_{\text{em}}$ - the photon soft factor in the case of there existing only multiple light particles in the final state
\begin{align}
\tilde{S}_{\text{em}} &= - \left(\frac{\epsilon^i}{\omega} \sum_a q_a \frac{1}{1 - \hat{n}.\vec{\beta}_a}  \beta_{a i}\right) - \left(i \ln \omega^{-1} \epsilon^i\sum_a q_a \left(C_a - 2 G M \right) \frac{1}{1 - \hat{n}.\vec{\beta}_a} \beta_{a i}\right) + \text{finite}
\label{ai.msoft}
\end{align}

The expression for $C_a$, for an individual outgoing light particle, follows from the expression for $C_+$ in Eq.~(\ref{c.pp}) with `$+$' replaced with `$a$'
\begin{equation}
C_a = {G M\over|\vec\beta_{a}|^3} (3\vec\beta_{a}^2-1) - \frac{q_a Q}{4\pi m_a  |\vec\beta_{a}|^3}(1-\vec\beta_{a}^2)^{3/2}\,.
\label{ca.pp}
\end{equation}

Following the analysis of~\cite{Laddha:2018vbn}, we now substitute Eq.~(\ref{ca.pp}) in Eq.~(\ref{eij.msoft}) and Eq.~(\ref{ai.msoft}) to express the graviton and photon soft factors in the following way
\begin{align}
\tilde{S}_{\text{gr}} &= i \epsilon^{ij}\left[i \omega^{-1} \tilde{A}_{ij}^{\text{gr}} + \ln \omega^{-1} \tilde{B}_{ij}^{\text{gr}}\right]  + \text{finite} \,, \notag\\
\tilde{S}_{\text{em}} &=  i \epsilon^i\left[i \omega^{-1} \tilde{A}_{i}^{\text{em}} + \ln \omega^{-1} \tilde{B}_{i}^{\text{em}}\right] + \text{finite} \,,
\label{mt.exp}
\end{align}
where 
\begin{align}
\tilde{A}_{ij}^{\text{gr}} &=  8 \pi G \sum_{a} m_a \frac{1}{1 - \hat{n}.\vec{\beta}_a}\frac{1}{\sqrt{1 - \vec{\beta}^2_a}}\beta_{ai} \beta_{aj} \,, \label{g.mem}\\
\tilde{B}_{ij}^{\text{gr}} &= 8 \pi G \sum_{a} \frac{1}{1- \hat{n}.\vec\beta_{a}}\frac{1}{ \sqrt{1- \vec\beta_{a}^2}}\frac{1}{\vert \vec\beta_{a} \vert^3}\left[ G M m_a \left(1- 3\vec\beta_{a}^2 + 2 \vert \vec\beta_{a} \vert^3\right) + \frac{Q q_a}{4 \pi}\left(1- \vec\beta_{a}^2\right)^\frac{3}{2}\right]\beta_{a i}\beta_{a j}\label{g.tail}\\
\tilde{A}_{i}^{\text{em}} &=  \sum_{a} q_a \frac{1}{1 - \hat{n}.\vec{\beta}_a}\beta_{ai} \,, \label{e.mem}\\
\tilde{B}_{i}^{\text{em}} &= \sum_{a} \frac{1}{1- \hat{n}.\vec\beta_{a}}\frac{1}{\vert \vec\beta_{a} \vert^3}\left[ G M q_a \left(1- 3\vec\beta_{a}^2 + 2 \vert \vec\beta_{a} \vert^3\right) + \frac{Q q^2_a}{4 \pi m_a}\left(1- \vec\beta_{a}^2\right)^\frac{3}{2}\right]\beta_{a i}\label{e.tail}
\end{align}
Ignoring the $Q$ dependent term in Eq.~(\ref{g.tail}), we see that the expressions in Eq.~(\ref{g.mem}) and Eq.~(\ref{g.tail}) agree with Eq.~(\ref{sl.mem}) and Eq.~(\ref{sl.tail}) respectively up to a factor of $\left(4 \pi R\right)^{-1}$. Hence $\tilde{A}^{\text{gr}}_{ij}$ represents the memory effect while $\tilde{B}_{ij}^{\text{gr}}$ represents the corresponding tail effect of late time gravitational waves on the RN spacetime. We can similarly take $\tilde{A}_{i}^{\text{em}}$ to represent the memory effect and $\tilde{B}_{i}^{\text{em}}$ to represent the tail effect of late time electromagnetic waveforms. The results on the Schwarzschild spacetime follow from setting $Q = 0$.

The $\tilde{B}^{\text{gr}}_{ij}$ and $\tilde{B}^{\text{em}}_{i}$ are tails of the memory term resulting as a consequence of logarithmic and phase contributions in the subleading soft factor. In considering the $\vert \vec\beta_a\vert \to 1$ limit we see that both $\tilde{B}_{ij}^{\text{gr}} \to 0$ and $\tilde{B}^{\text{em}}_{i} \to 0$. This implies a vanishing tail in both electromagnetic and gravitational waves in the ultrarelativistic limit. Thus in the case of only massless or ultrarelativistic outgoing particles, we only have the leading memory contribution. The vanishing tail at late times is a consequence of the phase contribution to the soft factor.

For $\vert \vec\beta_a\vert < 1$, the tail effect is present. If all masses and charges are comparable, we note from the dependence on $G$ in Eq.~(\ref{g.tail}) and Eq.~(\ref{e.tail}) that the charge $Q$ contribution will generically dominate that the mass $M$ contribution within the tail terms of gravitational and electromagnetic waves.  We can also determine from the integrals evaluated in the previous subsection that the expressions for $e_{ij}\left(t\,,\vec{x}\right)$ and $a_i\left(t\,,\vec{x}\right)$ in position space which provide Eq.~(\ref{mt.exp}) will take the form
\begin{align}
e_{ij}\left(t\,,\vec{x}\right) & \sim \tilde{A}^{\text{gr}}_{ij} + \frac{1}{t}\tilde{B}^{\text{gr}}_{ij} + \mathcal{O}\left(t^{-2}\right) \notag\\
a_{i}\left(t\,,\vec{x}\right) & \sim \tilde{A}^{\text{em}}_{i} + \frac{1}{t}\tilde{B}^{\text{em}}_{i} + \mathcal{O}\left(t^{-2}\right) \,.
\end{align}
Hence at late times, the leading memory effect will dominate the tail in both gravitational and electromagnetic waveforms. 

\section{Summary and Discussion} \label{disc}

We derived the electromagnetic and gravitational bremmstrahlung resulting from the classical scattering of a probe particle on the RN spacetime. The results were derived up to leading order in $\frac{M}{r}$ and $\frac{Q}{r}$, consistent with our assumption of a large impact parameter for the scattering.
Our results in Eq.~(\ref{hij.fin}) and Eq.~(\ref{ai.fin}) for the gravitational and electromagnetic cases respectively, can be noted as providing central charge $Q$ corrections to the known results for the gravitational radiation resulting from the probe mass scattering~\cite{Peters:1970mx} and the electromagnetic radiation resulting from the probe charge scattering~\cite{Peters:1973ah} on the Schwarzschild spacetime. The terms involving $Q$ in addition capture the coupling of gravitational and electromagnetic perturbations on the RN spacetime. This is noted through the specific contributions of $Q q$ in the gravitational solution Eq.~(\ref{hij.fin}) and $Q m$ in the electromagnetic solution Eq.~(\ref{ai.fin}).

The classical limit of soft theorems can be taken for scattering processes either involving large impact parameters or the probe-scatterer approximation~\cite{Laddha:2018rle}. Accordingly, in Sec.~\ref{soft} we applied the formalism of~\cite{Laddha:2018rle,Laddha:2018myi} to determine the soft factors in the probe-scatterer approximation, which includes the subleading phase corrections first described in~\cite{Sahoo:2018lxl} on the RN spacetime. Our result for the graviton soft factor in Eq.~(\ref{eij.soft}) and the photon soft factor in Eq.~(\ref{ai.soft1}) agrees with the predicted probe-scatterer approximation expression in the presence of gravitational and electromagnetic interactions.  In Sec.~\ref{gravemwav}, following the approach of~\cite{Laddha:2018vbn}, we used the soft factor expressions to determine the late time radiation from scattering processes involving no incoming and several outgoing light particles. This extension in particular covers merger events, where the energy of the ejected light particles are less than the rest mass of the central (merged) scatterer. The memory effect in four dimensional spacetimes involves a tail contribution arising from the logarithmic terms and phase corrections present in the subleading soft factor. We expressed the photon and graviton soft factors derived in Eq.~(\ref{ai.soft1}) and Eq.~(\ref{eij.soft}) respectively, in terms of the memory and tail effect contributions in Eq.~(\ref{mt.exp}). In all cases, the leading memory effect always dominates the tail contribution at late times, owing to the $t^{-1}$ fall-off of the tail term in the radiation.  

\bigskip

{\bf Acknowledgment:}
We thank Alok Laddha, Ashoke Sen, Biswajit Sahoo and Arnab Priya Saha for valuable discussions and feedback during the course of this work. KF would like to thank the CMI Chennai for their hospitality during the final stage of this work. KF was partially supported by the Max Planck Partner group “Quantum Black Holes” between CMI Chennai and AEI Potsdam and by a grant to CMI from the Infosys Foundation.

\appendix
\section{Scalar Green function on linearized curved spacetimes} \label{gfrn}
In flat spacetime, the scalar Green function satisfies the equation
\begin{equation}
\Box G\left(x,r(\sigma)\right) =\eta^{\mu\nu}\partial_{\mu}\partial_{\nu} G\left(x,r(\sigma)\right) = -\delta^{4}\left(x-r(\sigma)\right)
\end{equation}
which has the general solution
\begin{equation}
G\left(x,r(\sigma)\right)  = \frac{1}{4\pi} \delta\left(-\Omega_0\left(x,r(\sigma)\right)\right)\,,
\label{green.flat}
\end{equation}
where
\begin{align}
\Omega_0\left(x,r(\sigma)\right) &= \frac{1}{2}\eta_{\mu \nu}\left(x^{\mu} - r^{\mu}(\sigma)\right)\left(x^{\nu} - r^{\nu}(\sigma)\right) = \frac{1}{2} \left(-\left(t-r^0(\sigma)\right)^2 + \left(\vec{x} - \vec{r}(\sigma)\right)^2\right)\,.
\end{align}
We may further express Eq.~(\ref{green.flat}) as a sum over retarded and advanced Green functions
\begin{align}
G(x,r(\sigma)) =& \,G_R(x,r(\sigma)) + G_A(x,r(\sigma)) \notag\\
G_R(x,r(\sigma)) = \frac{1}{4\pi R_0} \delta\left(t - r^0(\sigma) - R_0 \right) \,,& \qquad  G_A(x,z(s)) = \frac{1}{4\pi r} \delta\left(t - r^0(\sigma) + R_0 \right)
\end{align}
where  
\begin{equation}
\vec{R}_0 = \vec{x} - \vec{r}(\sigma) \,, \qquad R_0 = \vert \vec{R}_0 \vert\,.
\label{R0}
\end{equation}
Hence Eq.~(\ref{scalar.cov}) in flat spacetime 
\begin{equation}
\Box \psi(x) = - \int \limits_{-\infty}^{\infty} \delta^{4}\left(x-r(\sigma)\right) f(\sigma) d\sigma
\label{flat.green}
\end{equation}
can have solutions involving only the retarded Green function by specifying limits on the integral of the full solution
\begin{align}
\psi(x) &=  \frac{1}{4\pi}\int \limits_{-\infty}^{\sigma_0} \delta\left(-\Omega_0\left(x,r(\sigma)\right)\right) f(\sigma) d\sigma = \frac{1}{4\pi} \int \limits_{-\infty}^{\infty} \frac{\delta\left(t - r^0(\sigma) - R_0 \right)}{R_0} f(\sigma) d\sigma \,,
\label{sol.flat}
\end{align}
where $\sigma_0$ is chosen such that $r^{\mu}(\sigma_0)$ lies outside the light cone centered on $x$. Analogous to $\Omega_0\left(x,r(\sigma)\right)$ in flat spacetime, we can define a world function $\Omega\left(x,r(\sigma)\right)$ on curved spacetimes which makes use of the geodesic distance between $x$ and $r(\sigma)$. Let $\xi^{\mu}(u)$ be the parametric solution of an unique geodesic connecting the points $x$ and $r(\sigma)$ which satisfies
\begin{equation}
\frac{d U^{\alpha}}{d u} + \Gamma^{\alpha}_{\mu\nu} U^{\mu}U^{\nu} = 0 \,,
\label{geod.eqn}
\end{equation}
where $U^{\alpha} = \frac{d\xi^{\alpha}}{d u}$ and the points $x$ and $r(\sigma)$ are taken to be at the parametric values of $u_1$ and $u_0$ respectively. We then define the world function in the following way
\begin{equation}
\Omega\left(x,r(\sigma)\right) = \frac{u_1 - u_0}{2} \int \limits_{u_0}^{u_1}  g_{\mu \nu}U^{\mu} U^{\nu}\, du \equiv \frac{\left(u_1 - u_0\right)^2}{2} g_{\mu \nu}U^{\mu} U^{\nu} \,,
\label{wf.u}
\end{equation}
where the last equality follows from the fact that since $U^{\mu}$ satisfies Eq.~(\ref{geod.eqn}), the integrand in the second equality of Eq.~(\ref{wf.u}) is constant. From Eq.~(\ref{wf.u}) we can derive the following relations 
\begin{equation}
\Omega\left(x,r(\sigma)\right)_{;\mu} = \left(u_1 - u_0\right) U_{\mu} \,, \qquad \Omega_{;\mu}\Omega_{;}{}^{\mu} = 2 \Omega \,.
\label{wf.id1}
\end{equation}
From considering the second derivatives of $\Omega$, we can also derive the following expression
\begin{equation}
\Omega\left(x,r(\sigma)\right)_{;\alpha}{}^{\alpha} = 4 - F\left(x,r(\sigma)\right) + \mathcal{O}\left(R^2\right) \,,
\label{wf.id2}
\end{equation}
where
\begin{equation}
F\left(x,r(\sigma)\right) = \frac{1}{u_1 - u_0} \int \limits_{u_0}^{u_1} \left(u-u_0\right)^2 R_{\mu \nu} U^{\mu} U^{\nu} du\,
\end{equation}
and $\mathcal{O}\left(R^2\right)$ represent terms which are quadratic and higher in curvature. As we will be seeking solutions about linearized curved backgrounds, we can ignore the $\mathcal{O}\left(R^2\right)$ contributions resulting from derivatives of the world function.
We also note that an equivalent definition of the world function can be provided in terms of the path length
\begin{equation}
\Omega\left(x,r(\sigma)\right) = -\frac{1}{2}\Delta S^2\left(x,r(\sigma)\right) \label{dist.exp}\,,
\end{equation}
where $\Delta S\left(x,r(\sigma)\right)$ is the integral over the geodesic distance (say $s$) from $x$ to $r(\sigma)$
\begin{equation}
\Delta S\left(x,r(\sigma)\right) = \int \limits_{r^0}^{t} \frac{ds}{d\xi^0}d\xi^0 \label{dist.int}\,.
\end{equation}
Using the world function, we first assume a trial solution of Eq.~(\ref{scalar.cov}) to be of the following form
\begin{align}
\psi^{(0)}(x) &=  \frac{1}{4\pi}\int \limits_{-\infty}^{\sigma_0} \delta\left(-\Omega\left(x,r(\sigma)\right)\right) f(\sigma) d\sigma \,.
\label{psi0}
\end{align}
Using Eq.~(\ref{wf.id1}) and Eq.~(\ref{wf.id2}), we then find from Eq.~(\ref{psi0}
\begin{equation}
\psi^{(0)}_{\phantom{(0)};\alpha}{}^{\alpha} = \frac{1}{4 \pi}\int \limits_{-\infty}^{\sigma_0} \left[2 \delta''\left(-\Omega\right) \Omega - 4 \delta'\left(-\Omega\right) + \delta'\left(-\Omega\right) F\left(x,r\right)\right] f(\sigma) d\sigma + \mathcal{O}\left(R^2\right)\,,
\end{equation}
where primes denote differentiation with respect to $-\Omega$. Re-expressing these derivatives as
\begin{equation}
\delta'\left(-\Omega\right) = \frac{d \delta\left(-\Omega\right)}{d \sigma} \frac{d\sigma}{d\left(-\Omega\right)}\,,
\end{equation}
we find
\begin{equation}
\psi^{(0)}_{\phantom{(0)};\alpha}{}^{\alpha} = - \int \limits_{-\infty}^{\infty} \delta^{4}\left(x-r(\sigma)\right) f(\sigma) d\sigma + \frac{1}{4 \pi}\int \limits_{-\infty}^{\sigma_0} \delta'\left(-\Omega\right) F(x,r(\sigma)) f(\sigma) d\sigma + \mathcal{O}\left(R^2\right)
\label{psi0.temp}
\end{equation}
From Eq.~(\ref{psi0.temp}), we can identify the following solution valid up to linear order in curvature
\begin{align}
\psi^{(1)}(x) &= \psi^{(0)}(x) + \delta \psi^{(0)}(x) \notag\\
&= \psi^{(0)}(x) + \frac{1}{16 \pi^2} \int \sqrt{-g(y)} \delta\left(-\Omega\left(x,y\right)\right) d^4y  \int \limits_{-\infty}^{\sigma_0} \delta'\left(-\Omega\left(y,r(\sigma)\right)\right) F\left(y,r(\sigma)\right) f(\sigma) d\sigma\,.
\label{psi1RN.sol}
\end{align}
Hence $\psi^{(1)}(x)$ in Eq.~(\ref{psi1RN.sol}) does satisfy Eq.~(\ref{scalar.cov}).

\section{Derivation of Eq.~(\ref{psiRN0.sol}) and Eq.~(\ref{dp.sol})} \label{app1}

\subsection{Derivation of Eq.~(\ref{psiRN0.sol})}

The $\psi^{(0)}$ solution given in Eq.~(\ref{psi0}) follows from the expression for the world function $\Omega\left(x\,, r\left(\sigma\right)\right)$ and properties of the delta function. 

We will derive the world function using Eq.~(\ref{dist.exp}) which involves the integral $\Delta S$ over the geodesic distance $s$ from $x$ to $r(\sigma)$. From Eq.~(\ref{dist.int}), we note that this further requires us to find the expression for $\frac{ds}{d\xi^0}$. This can be determined from the geodesic equation
\begin{equation}
 \frac{d^2 \xi^0}{ds^2} + \Gamma^0_{\alpha \beta}\frac{d \xi^{\alpha}}{ds} \frac{d \xi^{\beta}}{ds} = 0
\label{ge.eqn}
\end{equation}
Using the expression for $\Gamma^0_{0 k} = \phi_{,k}$, with $\phi \equiv \phi\left(\vec{\xi}\right)$ along the geodesic connecting $x$ with $r\left(\sigma\right)$, we then have from Eq.~(\ref{ge.eqn})
\begin{align}
&\frac{d^2 \xi^0}{ds^2} + 2 \frac{d \phi}{ds}  \frac{d \xi^{0}}{ds} = 0 \notag\\
& \Rightarrow \frac{d \xi^{0}}{ds} = A \exp\left[-2 \phi\right] \approx A\left(1 - 2 \phi\right) + \mathcal{O}\left(\phi^2\right)
\label{dxids}
\end{align}
Using the last line of Eq.~(\ref{dxids}) in Eq.~(\ref{dist.int}), we find
\begin{align}
\Delta S\left(x,r(\sigma)\right) = \int \limits_{r^0}^{t} \frac{ds}{d\xi^0}d\xi^0 &\approx \int \limits_{r^0}^{t} \frac{1}{A} \left(1 + 2 \phi\right)d\xi^0  \notag\\
& = \frac{t - r^0}{A} + \frac{2}{A} \int \limits_{r^0}^{t} \phi\left(\vec{\xi}\right) d \xi^0 \label{dist.int2}
\end{align}

We will solve the integral involved in the second line of Eq.~(\ref{dist.int2}) by first parametrizing the geodesic path in the following way 
\begin{equation}
\xi^{\mu} = r^{\mu} + \lambda \left( x^{\mu} - r^{\mu}\right)\,,
\label{gparam.eq}
\end{equation}
which is valid at the linearized the level. Hence
\begin{align}
d \xi^0  &= \left(t - r^0\right) d \lambda  \,, \label{gdparam.eq}\\
\phi\left(\vec{\xi}\right) &= - \frac{M}{8 \pi \vert \vec{r} - \lambda \vec{R}_0\vert} = -\frac{M}{ 8 \pi \sqrt{\vert \vec{r} \vert^2 + 2 \lambda R_0 \vert \vec{r} \vert \cos \theta + \lambda^2 R_0^2}} \,,
\label{sub.int2}
\end{align}
where $\vec{R}_0$ is as defined in Eq.~(\ref{R0}) and $\cos \theta = \frac{\vec{r}. \vec{R}_0}{\vert \vec{r}\vert R_0}$. Substituting Eq.~(\ref{gdparam.eq}) and Eq.~(\ref{sub.int2}) into the integral in the second line of Eq.~(\ref{dist.int2}), we have
\begin{align}
\frac{2}{A} \int \limits_{r^0}^{t} \phi\left(\vec{\xi}\right) d \xi^0 & = -\frac{M \left(t - r^0\right)}{4 \pi A} \int \limits_{0}^{1} \frac{1}{\sqrt{\vert \vec{r} \vert^2 + 2 \lambda R_0 \vert \vec{r} \vert \cos \theta + \lambda^2 R_0^2}} d \lambda \notag\\
&= -\frac{M \left(t - r^0\right)}{4 \pi R_0 A} \ln \left(\frac{\vec{x}.\vec{R}_0 + R R_0}{\vec{r}.\vec{R}_0 + \vert \vec{r}\vert R_0}\right) = -\frac{M \left(t - r^0\right)}{4 \pi R_0 A}\Gamma\left(\vec{x},\vec{r}\right)
\label{phi.int}
\end{align}
where we have replaced $\vert \vec{x} \vert = R$ and made use of the definition of $\Gamma\left(\vec{x},\vec{r}\right)$ as given in Eq.~(\ref{gamrho.exp}).
Using Eq.~(\ref{dist.int2}) and Eq.~(\ref{phi.int}) in Eq.~(\ref{dist.exp}), we then find
\begin{align}
\Omega\left(x,r\right) &= -\frac{1}{2}\Delta S^2\left(x,r\right) \notag\\
&= -\frac{\left(t - r^0\right)^2}{2 A^2} \left(1 -\frac{M}{2 \pi R_0} \Gamma\left(\vec{x},\vec{r}\right) \right) + \mathcal{O}\left(\phi^2\right)
\label{ds2.exp}
\end{align}
To complete our expression for $\Omega\left(x,r\right)$ we need to determine the constant $A^{-2}$, which follows from the normalization condition
\begin{equation}
g_{\mu \nu}\frac{d \xi^{\mu}}{ds} \frac{d \xi^{\nu}}{ds} = -1
\label{geod.norm}
\end{equation}
Denoting $\frac{d \xi^i}{d \xi^0} = V^i$, Eq.~(\ref{geod.norm}) can be re-expressed as
\begin{equation}
A^{-2} \left(1+ 4 \phi \right) = \left(1 + 2\phi\right) - \vert \vec{V}\vert^2 \left(1 - 2\phi\right)\,.
\label{vva.eq}
\end{equation}
We fix the constant by considering the asymptotic limit for large values of $\vert \vec{\xi}\vert$ and with $\phi \to 0$. In this case Eq.~(\ref{vva.eq}) simplifies to $A^{-2} = 1 - \vert \vec{V}_A\vert^2 $, where in denoting $\vert \vec{V} \vert$ by $\vert \vec{V}_A \vert$ we take this velocity to be the asymptotic value. Replacing this expression for $A^{-2}$ in Eq.~(\ref{vva.eq}) we can then find the first-order in $\phi$ corrections for $\vert \vec{V}\vert$. We can also take all $\phi \vert \vec{V}\vert^2 \sim \phi$, i.e. $\vert \vec{V}\vert^2 \approx c^2 = 1$ in all terms multiplying $\phi$, since all deviations are $\mathcal{O}\left(\phi^2\right)$. Accordingly, from Eq.~(\ref{vva.eq}) we then find $\vert \vec{V}\vert^2 = \vert \vec{V}_A\vert^2 + 4 \phi$ and hence    
\begin{equation}
\vert \vec{V}\vert = \vert \vec{V}_A\vert + 2 \phi + \mathcal{O}\left(\phi^2\right) \,. \label{V.exp}
\end{equation}
The path distance $\vert \vec{x} - \vec{r}\left(\sigma\right)\vert = R_0$ satisfies
\begin{equation}
R_0 = \int \limits_{r^0}^t \vert \vec{V}\left(\xi^0\right) \vert d \xi^0 \approx  \vert \vec{V}_A \vert \left(t - r^0\right) + 2 \int \limits_{r^0}^t \phi\left(\vec{\xi}\right) d \xi^0 \,,
\end{equation}
where we made use of Eq.~(\ref{V.exp}). Thus $\displaystyle{\vert \vec{V}_A \vert \approx \frac{1}{t - r^0} \left(R_0 - 2 \int \limits_{r^0}^t \phi\left(\vec{\xi}\right) d \xi^0\right)}$, which upon substituting in $A^{-2} = \left(1 -  \vert \vec{V}_A \vert^2 \right)$ gives
\begin{align}
A^{-2} &= \frac{1}{\left(t - r^0\right)^2}\left(\left(t - r^0\right)^2 - R_0^2 + 4 R_0 \int \limits_{r^0}^t \phi\left(\vec{\xi}\right) d \xi^0\right) + \mathcal{O}\left(\phi^2\right) \notag\\
& = \frac{1}{\left(t - r^0\right)^2}\left(\left(t - r^0\right)^2 - R_0^2 -  \frac{M R_0}{2 \pi} \Gamma\left(\vec{x},\vec{r}\right)\right) + \mathcal{O}\left(\phi^2\right)  \,,
\label{a.exp}
\end{align}
where we made use of Eq.~(\ref{phi.int}) and the property $M \frac{t - r^0}{R_0} = \frac{M}{\vert \vec{V}_A \vert} + \mathcal{O}\left(\phi^2\right) \sim M+ \mathcal{O}\left(\phi^2\right)$. Substituting Eq.~(\ref{a.exp}) in Eq.~(\ref{ds2.exp}), we find
\begin{align}
\Omega\left(x,r\right) &\approx -\frac{1}{2} \left(\left(t - r^0\right)^2 - R_0^2 -  \frac{M R_0}{2 \pi} \Gamma\left(\vec{x},\vec{r}\right)\right)\left(1 -\frac{M}{2 \pi R_0} \Gamma\left(\vec{x},\vec{r}\right) \right) \notag\\
& \approx -\frac{1}{2} \left(\left(t - r^0\right)^2 - \left(R_0 + \frac{M}{4 \pi} \Gamma\left(\vec{x},\vec{r}\right) \right)^2\right)\left(1 -\frac{M}{4 \pi R_0} \Gamma\left(\vec{x},\vec{r}\right) \right)^2 \,,
\label{om.exp}
\end{align}
where in the last line we could complete the squares since $\mathcal{O}\left(M^2\right)$ terms are ignored. 
Using the delta function scaling property
\begin{equation}
\delta\left(\frac{y}{a}\right) = a \delta\left(y\right)\,,
\label{delt.scal}
\end{equation}
where $a$ is a constant and the transformation property
\begin{equation}
\delta\left(f\left(y\right)\right) = \sum_{y_0} \frac{1}{\vert f'\left(y_0\right)\vert}\delta\left(y - y_0\right) \,,
\label{delt.tran}
\end{equation}
where $y_0$ is a non-repeated root of $f(y)$, we find on substituting Eq.~(\ref{om.exp}) in Eq.~(\ref{psi0}) 
\begin{align}
\psi^{(0)}(t\,,\vec{x}) &=  \frac{1}{4\pi}\int \limits_{-\infty}^{\sigma_0} \delta\left(-\Omega\left(x,r(\sigma)\right)\right) f(\sigma) d\sigma \notag\\
& = \frac{1}{4 \pi} \int \limits_{-\infty} ^{\infty} \frac{\delta\left(t - r^0 - R_0 -  \frac{M}{4\pi} \Gamma\left(\vec{x},\vec{z}\right) \right)}{R_0} f(\sigma) d\sigma + \mathcal{O}(M^2)
\label{psi0sol.der}
\end{align}
The advanced Green function contribution is absent due to the upper limit on the integral in the first line of Eq.~(\ref{psi0sol.der}).

\subsection{Derivation of Eq.~(\ref{dp.sol})}

In the case of $\delta \psi^{(0)}$, we find the following equation from Eq.~(\ref{psi1.sol})
\begin{align}
\delta \psi^{(0)}(x)= \frac{1}{16 \pi^2} \int \sqrt{-g(y)} \delta\left(-\Omega\left(x,y\right)\right) d^4y  \int \limits_{-\infty}^{\sigma_0} \delta'\left(-\Omega\left(y,r(\sigma)\right)\right) F\left(y,r(\sigma)\right) f(\sigma) d\sigma\,,
\label{dpsi.eq}
\end{align}
with
\begin{equation}
F\left(y ,r(\sigma)\right) = \frac{1}{u_1 - u_0} \int \limits_{u_0}^{u_1} \left(u-u_0\right)^2 R_{\mu \nu} U^{\mu} U^{\nu} du\,
\label{f.app}
\end{equation}
where $U^{\alpha} = \frac{d\xi^{\alpha}}{d u}$ and the points $y$ and $r(\sigma)$ are taken to be at the parametric values of $u_1$ and $u_0$ respectively.
Since we are working up to linear order in $\phi$ and the distance of $r\left(\sigma\right)$ to black hole is taken to be large, we can approximate the black hole as a massive and charged point particle and replace $R_{\mu \nu}$ by its stress tensor. In other words, we consider
\begin{equation}
R_{\mu\nu} \approx M \delta^{(3)}\left(\vec{r}\,'\right) \delta^0_{\mu}  \delta^0_{\nu} + \mathcal{O}\left(\phi^2\right) \,,
\label{rbh.app}
\end{equation}
where $\vec{r}\,'$ denotes the location of the black hole and $\mathcal{O}\left(\phi^2\right)$ involve the $\mathcal{O}\left(Q^2\right)$ terms in the electromagnetic stress tensor. The evaluation of Eq.~(\ref{dpsi.eq}) also requires us to have a parametric expression for $\vec{r}\,'$ in terms of $u$. For this, we consider $U^{\alpha}$ to be nearly null, i.e.
\begin{equation}
\left(U^0\right)^2 \approx  \vec{U}^2 = \left(\frac{\vec{y} - \vec{r}}{u_1 - u_0}\right)^2 \,,
\label{u2.app}
\end{equation}
with the location of $\vec{r}\,'$ determined from the parametric relation for a straight-line path 
\begin{equation}
\vec{r}\,'\left(u\right) = \vec{r} + \left(u - u_0\right) \vec{U} = \vec{r} + \left(\frac{u - u_0}{u_1 - u_0}\right) \left(\vec{y} - \vec{r}\right)\,.
\label{rp.app}
\end{equation}
Substituting Eq.~(\ref{rbh.app}), Eq.~(\ref{u2.app}) and Eq.~(\ref{rp.app}) in Eq.~(\ref{f.app}), we find
\begin{align}
F\left(y ,r \right) &=  \frac{M}{u_1 - u_0} \int \limits_{u_0}^{u_1} \left(u-u_0\right)^2 \delta^{(3)}\left(\vec{r}\,'(u)\right) \left(U^0\right)^2 du + \mathcal{O}\left(\phi^2\right) \notag\\
&=  \frac{M}{\left(u_1 - u_0\right)^3} \int \limits_{u_0}^{u_1} \left(u-u_0\right)^2 \delta^{(3)}\left(\vec{r}\,'(u)\right) \left(\vec{y} - \vec{r}\right)^2 du + \mathcal{O}\left(\phi^2\right) \notag\\
&=  M \int \limits_{u_0}^{u_1} \frac{1}{u-u_0} \delta^{(3)}\left(\vec{y} + \vec{r} \left(\frac{u_1 - u}{u - u_0}\right) \right) \left(\vec{y} - \vec{r}\right)^2 du + \mathcal{O}\left(\phi^2\right)
\label{f.sub}
\end{align}
where we made use of the delta function scaling property Eq.~(\ref{delt.scal}) in the last line of Eq.~(\ref{f.sub}). 
For the terms other than $F$ involved in the integrand of Eq.~(\ref{dpsi.eq}), we can use the flat spacetime expressions since Eq.~(\ref{f.sub}) is $\mathcal{O}\left(M\right)$. Thus 
\begin{align}
\int \sqrt{-g(y)} \delta\left(-\Omega\left(x,y\right)\right) d^4y &\to \int \delta\left(-\Omega_0\left(x,y\right)\right) d^4y   \notag\\
&= \int\limits_{-\infty}^{\infty} \frac{1}{\vert \vec{x} - \vec{y}\vert}\delta\left(t - y^0 - \vert \vec{x} - \vec{y}\vert \right) d^4y \,,
\label{do.app}
\end{align}
where we made use of Eq.~(\ref{sol.flat}) in expressing the result as an integral over the retarded Green function.
We also change the derivative involved in the $\delta'\left(-\Omega\left(y,r(\sigma)\right)\right)$ term so that we differentiate with respect to the argument involved in the retarded Green function
\begin{align}
\frac{d}{d\left(- \Omega_0\right)} &=  \frac{d \left(y^0 - r^0 - \vert \vec{y} - \vec{r}\vert\right)}{d\left(- \Omega_0\right)}\frac{d}{d \left(y^0 - r^0 - \vert \vec{y} - \vec{r}\vert\right)}\notag\\
& = \frac{2}{y^0 - r^0 + \vert \vec{y} - \vec{r}\vert}\frac{d}{d \left(y^0 - r^0 - \vert \vec{y} - \vec{r}\vert\right)}\,.
\label{der.id}
\end{align}
Using the equation in the second line of Eq.~(\ref{der.id}) and the property that the path from $y$ to $r\left(\sigma\right)$ is nearly null, we then have
\begin{equation}
\int \limits_{-\infty}^{\sigma_0} \delta'\left(-\Omega\left(y,r(\sigma)\right)\right) f(\sigma) d\sigma \to
\int \limits_{-\infty}^{\infty} \frac{1}{\vert \vec{y} - \vec{r} \vert^2}\delta'\left(y^0 - r^0 - \vert \vec{y} - \vec{r}\vert\right) f(\sigma) d\sigma\,.
\label{ddo.app}
\end{equation}
Unless stated otherwise, $\delta'(A)$ will always mean that we differentiate with respect to the argument of the delta function `$A$'. Substituting Eq.~(\ref{f.sub}), Eq.~(\ref{do.app}) and Eq.~(\ref{ddo.app}) in Eq.~(\ref{dpsi.eq}), we get
\begin{align}
\delta \psi^{(0)}(x)&= \frac{M}{16 \pi^2}\int\limits_{-\infty}^{\infty} \frac{\delta\left(t - y^0 - \vert \vec{x} - \vec{y}\vert \right)}{\vert \vec{x} - \vec{y}\vert} d^4y \int \limits_{-\infty}^{\infty} \delta'\left(y^0 - r^0 - \vert \vec{y} - \vec{r}\vert\right) f(\sigma) d\sigma \notag\\ & \qquad\qquad\qquad\qquad\qquad \int \limits_{u_0}^{u_1} \frac{1}{u-u_0} \delta^{(3)}\left(\vec{y} + \vec{r} \left(\frac{u_1 - u}{u - u_0}\right) \right) du
\label{psi0.app}
\end{align}
On evaluating the $d^3y$ integral and denoting $\alpha = \frac{u_1 - u}{u - u_0}$ in Eq.~(\ref{psi0.app}), we have
\begin{equation}
\delta \psi^{(0)}(x)= \frac{M}{16 \pi^2}\int\limits_{-\infty}^{\infty} \frac{\delta\left(t - y^0 - \vert \vec{x} + \vec{r} \alpha\vert \right)}{\vert \vec{x} + \vec{r} \alpha \vert} dy^0 \int \limits_{-\infty}^{\infty} \delta'\left(y^0 - r^0 - \vert \vec{r} (1 + \alpha)\vert\right) f(\sigma) d\sigma\int \limits_{u_0}^{u_1} \frac{d u}{u-u_0}
\label{psi0d3y.app}
\end{equation}
Changing the variable from $\vert \vec{r} \vert \alpha = v$ gives
\begin{equation}
\frac{du}{u-u_0} = -\frac{dv}{\vert\vec{r}\vert + v}
\end{equation}
Hence changing the variable from $u$ to $v$ in Eq.~(\ref{psi0d3y.app}), we find
\begin{equation}
\delta \psi^{(0)}(x)= \frac{M}{16 \pi^2} \int \limits_{0}^{\infty} dv \int \limits_{-\infty}^{\infty} \frac{\delta\left(t - y^0 - \rho(v)\right)}{\rho(v) \left(\vert \vec{r}\vert + v\right)} dy^0 \int \limits_{-\infty}^{\infty} \delta'\left(y^0 - r^0 - \vert \vec{r}\vert - v \right) f(\sigma) d\sigma\,,
\label{psi0v.app}
\end{equation}
where $\rho(v)$ is as defined in Eq.~(\ref{gamrho.exp}), i.e. 
\begin{equation}
\rho(v) = \sqrt{R^2 + v^2 + \frac{2 v \vec{x}.\vec{r}}{\vert \vec{r}\vert}} \,, \quad R = \vert \vec{x}\vert
\end{equation}

To derive Eq.~(\ref{dp.sol}) from Eq.~(\ref{psi0v.app}) we proceed by first making use of the shift property of delta function to get
\begin{align}
\delta \psi^{(0)}(x) &= \frac{M}{16 \pi^2} \int \limits_{0}^{\infty} dv \int \limits_{-\infty}^{\infty} \frac{\delta\left(t - y^0\right)}{\rho(v) \left(\vert \vec{r}\vert + v\right)} dy^0 \int \limits_{-\infty}^{\infty} \delta'\left(y^0 - r^0 - \vert \vec{r}\vert - v  - \rho(v) \right) f(\sigma) d\sigma \,.
\end{align}
We now change the derivative acting on the delta function in the following way
\begin{align}
\delta'\left(y^0 - r^0 - \vert \vec{r}\vert - v - \rho(v)\right) &= \frac{d y^0}{d\left(y^0 - r^0 - \vert \vec{r}\vert - v - \rho(v) \right)} \frac{d}{d y^0}\delta\left(y^0 - r^0 - \vert \vec{r}\vert - v - \rho(v) \right)\notag\\ &= \frac{d}{d y^0}\delta\left(y^0 - r^0 - \vert \vec{r}\vert - v - \rho(v) \right) \,.
\end{align}
Since $\rho(v) \left(\vert \vec{r}\vert + v\right)$ is independent of $y^0$, we thus have
\begin{equation}
\delta \psi^{(0)}(x) = \frac{M}{16 \pi^2} \int \limits_{-\infty}^{\infty} dy^0 \delta\left(t - y^0\right)  \frac{d}{d y^0}\int \limits_{0}^{\infty} dv \int \limits_{-\infty}^{\infty} \frac{\delta \left(y^0 - r^0 - \vert \vec{r}\vert - v  - \rho(v) \right)}{\rho(v) \left(\vert \vec{r}\vert + v\right)} f(\sigma) d\sigma
\label{psi0v1.app}
\end{equation}
Finally, by evaluating the $y^0$ integral, we get Eq.~(\ref{dp.sol}).

\end{document}